\documentclass[11pt]{article}
\usepackage{latexsym, a4wide}
\usepackage{amsmath, rotating, color}
\usepackage{amsfonts,amssymb, amsthm}
\usepackage{pictexwd}

\newcommand{\dickm}[1]{\text{\boldmath ${#1}$}}

\newtheorem{theorem}{Theorem}
\newtheorem{proposition}{Proposition}[section]

\newtheorem{definition}[proposition]{Definition}
\theoremstyle{definition}
\newtheorem{remark}[proposition]{Remark}
\newtheorem{example}[proposition]{Example}
\newtheoremstyle{step}{3pt}{0pt}{}{}{\bf}{}{.5em}{}
\theoremstyle{step} \newtheorem{step}{Step}

\numberwithin{equation}{section}

\newcommand\unnumberedfootnote[1]{ %
        \let\temp=\thefootnote %
        \renewcommand{\thefootnote}{}%
        \footnote{#1}%
        \let\thefootnote=\temp%
        \addtocounter{footnote}{-1}}

\newcommand{\one}[1]{\text{\parbox{1cm}{\beginpicture
\setcoordinatesystem units <.1cm,.1cm>
\setplotarea x from 0 to 6, y from 3 to 13
\plot 3 3 3 8 /
\put{\footnotesize#1} [cC] at 3 11
\multiput{\tiny $\bullet$} at 3 3 *50  0 0.1 /
\endpicture}}}

\newcommand{\Y}{\text{\parbox{1.5cm}{
\beginpicture
\setcoordinatesystem units <0.1cm, 0.1cm>
\setplotarea x from 0 to 20, y from 3 to 17
\plot 10 6 10 10 7 13 10 10 13 13 /
\put{\footnotesize$\pi'_{(j)}$} [cC] at 4.5 15
\put{\footnotesize$\pi'_{(k)}$} [cC] at 15 15 
\endpicture}}}

\newcommand{\Yup}{\text{\parbox{1.5cm}{
\beginpicture
\setcoordinatesystem units <0.1cm, 0.1cm>
\setplotarea x from 0 to 20, y from 3 to 17
\plot 10 6 10 10 7 13 10 10 13 13 /
\multiput{\tiny $\bullet$} at 10 10 *100  0.03 0.03 /
\multiput{\tiny $\bullet$} at 10 10 *100  -0.03 0.03 /
\put{\footnotesize$\pi'_{(j)}$} [cC] at 4.5 15
\put{\footnotesize$\pi'_{(k)}$} [cC] at 15 15 
\endpicture}}}

\newcommand{\Yri}{\text{\parbox{1.5cm}{
\beginpicture
\setcoordinatesystem units <0.1cm, 0.1cm>
\setplotarea x from 0 to 20, y from 3 to 17
\plot 10 6 10 10 7 13 10 10 13 13 /
\multiput{\tiny $\bullet$} at 10 10 *100  0.03 0.03 /
\multiput{\tiny $\bullet$} at 10 10 *100  0 -0.04 /
\put{\footnotesize$\pi'_{(j)}$} [cC] at 4.5 15
\put{\footnotesize$\pi'_{(k)}$} [cC] at 15 15 
\endpicture}}}

\newcommand{\Yle}{\text{\parbox{1.5cm}{
\beginpicture
\setcoordinatesystem units <0.1cm, 0.1cm>
\setplotarea x from 0 to 20, y from 3 to 17
\plot 10 6 10 10 7 13 10 10 13 13 /
\multiput{\tiny $\bullet$} at 10 10 *100  -0.03 0.03 /
\multiput{\tiny $\bullet$} at 10 10 *100  0 -0.04 /
\put{\footnotesize$\pi'_{(j)}$} [cC] at 4.5 15
\put{\footnotesize$\pi'_{(k)}$} [cC] at 15 15 
\endpicture}}}

\newcommand{\Ybottom}{\text{\parbox{1.5cm}{
\beginpicture
\setcoordinatesystem units <0.1cm, 0.1cm>
\setplotarea x from 0 to 20, y from 3 to 17
\plot 10 6 10 10 7 13 10 10 13 13 /
\multiput{\tiny $\bullet$} at 10 10 *100  0 -0.04 /
\put{\footnotesize$\pi'_{(j)}$} [cC] at 4.5 15
\put{\footnotesize$\pi'_{(k)}$} [cC] at 15 15 
\endpicture}}}

\newcommand{\Yupri}{\text{\parbox{1.5cm}{
\beginpicture
\setcoordinatesystem units <0.1cm, 0.1cm>
\setplotarea x from 0 to 20, y from 3 to 17
\plot 10 6 10 10 7 13 10 10 13 13 /
\multiput{\tiny $\bullet$} at 10 10 *100  0.03 0.03 /
\put{\footnotesize$\pi'_{(j)}$} [cC] at 4.5 15
\put{\footnotesize$\pi'_{(k)}$} [cC] at 15 15 
\endpicture}}}

\newcommand{\Yall}{\text{\parbox{1.5cm}{
\beginpicture
\setcoordinatesystem units <0.1cm, 0.1cm>
\setplotarea x from 0 to 20, y from 3 to 17
\plot 10 6 10 10 7 13 10 10 13 13 /
\multiput{\tiny $\bullet$} at 10 10 *100  0.03 0.03 /
\multiput{\tiny $\bullet$} at 10 10 *100  -0.03 0.03 /
\multiput{\tiny $\bullet$} at 10 10 *100  0 -0.04 /
\put{\footnotesize$\pi'_{(j)}$} [cC] at 4.5 15
\put{\footnotesize$\pi'_{(k)}$} [cC] at 15 15 
\endpicture}}}

\begin{document}
\title{\LARGE Approximating genealogies for partially linked neutral
  loci under a selective sweep}

\thispagestyle{empty}

\author{{\sc by P. Pfaffelhuber\thanks{Corresponding author; Tel: (+49)-89-74-2180-108; email: p.p@lmu.de} and A. Studeny} \\[2ex]
  \emph{Ludwig-Maximilian University Munich} \vspace*{-7ex}} \date{}

\maketitle
\unnumberedfootnote{\emph{AMS 2000 subject classification.} 92D15
  (Primary), 60J80, 60J85, 60K37, 92D10 (Secondary).}

\unnumberedfootnote{\emph{Keywords and phrases.} Selective sweep,
  genetic hitchhiking, diffusion approximation, Yule process,
  ancestral recombination graph, random background}

\begin{abstract}
\noindent
Consider a genetic locus carrying a strongly beneficial allele which
has recently fixed in a large population. As strongly beneficial
alleles fix quickly, sequence diversity at partially linked neutral
loci is reduced. This phenomenon is known as a \emph{selective sweep}.

The fixation of the beneficial allele not only affects sequence
diversity at single neutral loci but also the joint allele
distribution of several partially linked neutral loci. This
distribution can be studied using the ancestral recombination graph
for samples of partially linked neutral loci during the selective
sweep. To approximate this graph, we extend recent work by
\cite{SchweinsbergDurrett2005, EtheridgePfaffelhuberWakolbinger2006}
using a marked Yule tree for the genealogy at a single neutral locus
linked to a strongly beneficial one.

We focus on joint genealogies at two partially linked neutral loci in
the case of large selection coefficients $\alpha$ and recombination
rates $\rho=\mathcal O(\alpha/\log\alpha)$ between loci.  Our approach
leads to a full description of the genealogy with accuracy of
$\mathcal O((\log \alpha)^{-2})$ in probability. As an application, we
derive the expectation of Lewontin's $D$ as a measure for non-random
association of alleles.
\end{abstract}

\section{Introduction}
The model of \emph{selective sweeps}, also known as \emph{genetic
  hitchhiking}, introduced by Maynard-Smith and Haigh in
\cite{MaynardSmithHaigh1974}, is the starting point for a large body
of both empirical and theoretical population genetic studies
(\cite{Nurminsky2005}). It predicts that sequence diversity is reduced
close to a strongly selected locus on a recombining genome near the
time of fixation of the beneficial allele. Theoretical studies aim at
describing these patterns of genetic diversity in detail while
empirical work uses this prediction to identify genes under selection.

If a species or a population adapts to its environment, several genes
might be under strong selection. Moreover, if the function of genes
were known, we would have predictions as to which genes are
responsible for the adaptive process. Unfortunately, functional
information is scarce. Without functional knowledge and in the
presence of recombination, the model of selective sweeps helps to
identify candidate genes affected by recent selective pressures.
Genome scans are carried out for a sample of individuals, which show
patterns of sequence diversity at lots of marker loci in the whole
genome (\cite{NielsenEtAl2005}). If a marker shows low diversity,
statistical tests help to decide if a gene under selection is located
nearby (\cite{KimStephan2002, LiStephan2005}).

Most theoretical studies of selective sweeps have focused on a model
with one selective and one partially linked neutral locus
(\cite{MaynardSmithHaigh1974, StephanWieheLenz1992,
  KaplanHudsonLangley1989, Barton1998, SchweinsbergDurrett2005,
  EtheridgePfaffelhuberWakolbinger2006}). This simple model already
describes the reduction in sequence diversity. However genetic data
are frequently available for many partially linked loci. This raises
the question of whether selective sweeps also generate distinct
patterns of multi-locus allele frequencies. We will follow
\cite{StephanSongLangley2006} and study a three locus model with one
selective and two partially linked neutral loci. Using this model, it
is possible to study the non-random association of allelic types at
the two neutral loci, which is usually called \emph{linkage
  disequilibrium}.



An influential idea in the analysis of selective sweeps was to study
approximate genealogies describing relationships between the
individuals in a sample from the population. Studying genealogies at
the selected site started with \cite{KaplanDardenHudson1988} and was
carried further to linked neutral loci in
\cite{KaplanHudsonLangley1989}.

The genealogy at a single neutral locus can be constructed as a
\emph{structured coalescent}.  Here, the beneficial and wild-type
allele at the selected locus form two subpopulations. Their sizes are
determined by the frequency path of the beneficial allele during the
selective sweep. Assume a new gamete is built (forward in time) by
recombination of a beneficial allele at the selected locus and a
neutral variant linked to a wild-type. Following the neutral variant
backward in time leads to a migration event from the beneficial to the
wild-type background. Therefore, recombination acts as migration
between the beneficial and the wild-type backgrounds.

Genealogies of two or more loci can be constructed using the ancestral
recombination graph (\cite{Hudson1983, GriffithsMarjoram1997}).
Therefore, we will construct ancestries of two partially linked
neutral loci under a selective sweep by a \emph{structured ancestral
  recombination graph}.  As in the case of only one locus, the two
subpopulations are distinguished by the beneficial and wild-type
allele at the selected locus, respectively. This ancestral
recombination graph will serve as the exact model for genealogies at
partially linked loci under a selective sweep. However, an exact
analysis is hard to obtain, because the graph must be conditioned on
the random frequency path of the beneficial allele.

An alternative approach uses a two-step procedure for genealogies at
the selective and the neutral locus. First, the (approximate) genealogy at
the selective locus is generated and second, the genealogy at the
neutral locus is added, which might differ due to recombination.  Two
approximate genealogies at the selected site have been proposed.
First, a star-like genealogy, which means that the most recent common
ancestor of all pairs in the population is the individual which
carried the beneficial allele first (\cite{SchweinsbergDurrett2005,
  NielsenEtAl2005}). Second, a Yule process, i.e., a pure birth
process, which allows for coalescences also during the selective sweep
(\cite{SchweinsbergDurrett2005,
  EtheridgePfaffelhuberWakolbinger2006}). It was shown in
\cite[Theorems 1.1, 1.2]{SchweinsbergDurrett2005} that the Yule
process approximation is more exact than the star-like approximation.
Therefore, we will use this Yule process approximation for the
genealogy at the selected site to study the three locus model of
\cite{StephanSongLangley2006} for selective sweeps. We will show that
the analysis carried out in
\cite{EtheridgePfaffelhuberWakolbinger2006} in the two locus case can
be extended to the three locus case (Theorem \ref{T}). Moreover, the
approximation by a Yule process can be used to calculate
characteristics of linkage disequilibrium explicitly (Theorem
\ref{T2}).

\section{The model}
Consider a beneficial allele which enters a population of (haploid)
size $N$ at time $t=0$ and has a selective advantage of $s$ with
respect to the wild-type allele. Set $\alpha=sN$, which is called the
scaled selection coefficient. As selection can only be detected if the
beneficial allele fixes in the population, we condition on fixation of
the beneficial allele and let $T$ be the (random) time of fixation.

\smallskip

Assume reproduction in the population follows a Wright-Fisher model,
or, more generally, a Cannings model with individual offspring
variance 1. In the limit of infinite $N$ and a time rescaling in units
of $N$ generations, the frequency path of the beneficial allele is the
solution of the SDE
\begin{equation}\label{eq:SDE}
  dX = \alpha X(1-X)\coth(\alpha X)dt + \sqrt{X(1-X)}dW,
\end{equation}
with a standard Brownian motion $W$ and $X_0=0$. This diffusion arises
as $h$-transform of the process describing the unconditional frequency
path with the fixation probability of the beneficial allele as a
harmonic function and has $0$ as an entrance boundary. (See e.g.
\cite{Griffiths2003}, p. 245 and
\cite{EtheridgePfaffelhuberWakolbinger2006}, (2.1).)

\begin{figure}
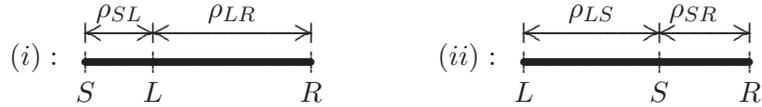

$$
(i): \text{
\parbox{2cm}{\beginpicture
\setcoordinatesystem units <1.5cm,1.3cm>
\setplotarea x from 0 to 2, y from 0 to 1
\plot 0 0.5 2 0.5 /
\plot 0 0.4 0 0.6 /
\plot 0.6 0.4 0.6 0.6 /
\plot 2 0.4 2 0.6 /
\put{$S$} [cC] at 0 0.2
\put{$L$} [cC] at 0.6 0.2
\put{$R$} [cC] at 2 0.2
\plot 0 0.7 0 0.9 /
\plot 0.6 0.7 0.6 0.9 /
\plot 2 0.7 2 0.9 /
\multiput {\tiny $\bullet$} at 0 .5 *200 .01 .0 /
\arrow <0.2cm> [0.375,1] from 0 .8 to .6 .8
\arrow <0.2cm> [0.375,1] from 0.6 .8 to 0 .8
\arrow <0.2cm> [0.375,1] from 2 .8 to .6 .8
\arrow <0.2cm> [0.375,1] from 0.6 .8 to 2 .8
\put{$\rho_{SL}$} [cC] at .3 1
\put{$\rho_{LR}$} [cC] at 1.3 1
\endpicture}} 
\qquad\qquad
(ii): \text{
\parbox{2cm}{\beginpicture
\setcoordinatesystem units <1.5cm,1.3cm>
\setplotarea x from 0 to 2, y from 0 to 1
\plot 0 0.5 2 0.5 /
\plot 0 0.4 0 0.6 /
\plot 1.2 0.4 1.2 0.6 /
\plot 2 0.4 2 0.6 /
\put{$L$} [cC] at 0 0.2
\put{$S$} [cC] at 1.2 0.2
\put{$R$} [cC] at 2 0.2
\plot 0 0.7 0 0.9 /
\plot 1.2 0.7 1.2 0.9 /
\plot 2 0.7 2 0.9 /
\multiput {\tiny $\bullet$} at 0 .5 *200 .01 .0 /
\arrow <0.2cm> [0.375,1] from 0 .8 to 1.2 .8
\arrow <0.2cm> [0.375,1] from 1.2 .8 to 0 .8
\arrow <0.2cm> [0.375,1] from 2 .8 to 1.2 .8
\arrow <0.2cm> [0.375,1] from 1.2 .8 to 2 .8
\put{$\rho_{LS}$} [cC] at .6 1
\put{$\rho_{SR}$} [cC] at 1.5 1
\endpicture}} 
$$
\caption{\label{fig:geom}The two possible geometries of the selected
  ($S$) and the two neutral loci ($L$ and $R$). The scaled
  recombination rates between loci are given by $\rho_{SL}, \rho{LR},
  \rho_{LS}$ and $\rho_{SR}$.}
\end{figure}

Two neutral loci are partially linked to the selected locus. For
simplicity, we refer to the two neutral loci as the \emph{l}eft and
\emph{r}ight neutral locus, denoted by $L$ and $R$. As illustrated in
Figure \ref{fig:geom}, the selected locus lies either (i) outside or
(ii) in between the neutral loci. All other possible geometries are
equivalent to either (i) or (ii) because of the symmetry in the model.

Recombination can break up the association of these three loci. (We
only consider recombination as simple crossing over. Gene conversion
is not considered in our model.) As we take a limiting infinite
population and rescale time by a factor of $N$, we have to consider
scaled recombination rates. These are different for the two
geometries. For geometry (i) we denote the recombination rates between
the selective and neutral loci by $\rho_{SL}$, $\rho_{LR}$ and for
geometry (ii) by $\rho_{LS}$, $\rho_{SR}$ respectively.

The two linked neutral loci do not affect the frequency path of the
beneficial allele. In contrast, neutral variants which are linked to
the beneficial allele at the beginning of the selective sweep rise in
frequency.  Looking backward in time from the time $T$ of fixation, we
can trace back the history of a finite sample at all three loci. As
the neutral loci are linked to the selected one, the genealogies at
all three loci are correlated.

For the construction of the ancestral recombination graph relating all
loci, time is running backward, so we set $\beta = T-t$. Conditioned
on a frequency path $\mathcal X=(X_t)_{0\leq t\leq T}$, given by
\eqref{eq:SDE}, we will describe the ancestral recombination graph as
a partition-valued process $\xi^{\mathcal X} =(\xi^{\mathcal
  X}_\beta)_{0\leq \beta\leq T}$.

\smallskip

Assume we take a sample from the population at time $T$. Every
individual in the sample carries one $L$ and one $R$-locus. Of all
$L$- and $R$-loci present in the sample we want to trace back a number
$\ell$ of $L$- and $r$ of $R$-loci. These loci are represented by sets
$\dickm\ell$ for the $L$- and $\dickm r$ for the $R$-loci. So, $\ell
:= |\dickm \ell|, r:= |\dickm r|$. To define the state space of the
structured ancestral recombination graph denote by $\mathcal P_A$ the
set of partitions of $A$ for a finite set $A$ and define
$$ \mathcal P'_{\dickm \ell\,\cup\,\dickm r} := \{\xi=(\xi^B, \xi^b), \xi^B\cup\xi^b \in\mathcal
P_{\dickm\ell \cup \dickm r}, \xi^B\cap\xi^b=\varnothing\}.$$ The
coordinates $\xi^B$ and $\xi^b$ contain partition elements located in
the beneficial and the wild-type background, respectively. For
$\xi\in\mathcal P'_{\dickm \ell \cup \dickm r}$ we write $\xi_{(j)}$
for the partition element containing $j\in\dickm \ell\cup \dickm r$.

The ancestral process is started at the time $\beta=0$ of fixation of
the beneficial allele. So, the sample of $L$- and $R$-loci is linked
to the beneficial allele. Therefore, we start the process in
$\xi^{\mathcal X}_0=(\pi, \varnothing)$ for some $\pi\in \mathcal
P_{\dickm \ell\,\cup\,\dickm r}$. Assume the state at time $\beta$ is
$\xi^{\mathcal X}_\beta=(\xi^B, \xi^b)\in\mathcal P'_{\dickm
  \ell\,\cup\,\dickm r}$. For $j\in\dickm \ell \cup \dickm r$ the
partition element which contains $j\in\dickm \ell$, i.e.,
$(\xi^{\mathcal X}_\beta)_{(j)}$, encodes the set of $L$- and
$R$-loci, taken from the population at time $T$, which have the same
ancestor as $j$ at time $T-\beta$.  Usually we will study the
genealogy of $n$ pairs of $L$- and $R$-loci. In this case set $\dickm
\ell:=\{1,\ldots, n\}$ and $\dickm r:=\{n+1,\ldots, 2n\}$ and start
the process with $\pi = \{\{1,n+1\}, \ldots, \{n,2n\}\}$.

The dynamics of the process is given as follows: Coalescence events
occur for lines in the beneficial and the wild-type background with
pair coalescence rate $1/X_{T-\beta}$ and $1/(1-X_{T-\beta})$ at time
$\beta$, respectively. So, given $\xi^{\mathcal
  X}_\beta=(\xi^B,\xi^b)$ with $\xi^B=\{\xi^B_1, \ldots,
\xi^B_{|\xi^B|}\}$ and $\xi^b=\{\xi^b_1, \ldots, \xi^b_{|\xi^b|}\}$
transitions occur for $1\leq j\neq k\leq |\xi^B|$ and $1\leq j'\neq
k'\leq |\xi^b|$ from $(\xi^B, \xi^b)$ to
\begin{equation}\label{eq:coal}
\begin{aligned}
  &\left((\xi^B\setminus \{\xi^B_j, \xi^B_k\}) \cup \{\xi_j^B\cup
    \xi^B_k\}, \xi^b)\right) &&
  \text{ with rate }\frac{1}{X_{T-\beta}},  & \quad(1)\\
  & \left((\xi^B,(\xi^b\setminus \{\xi^b_{j'}, \xi^b_{k'}\}) \cup
    \{\xi_{j'}^b\cup \xi^b_{k'}\})\right) && \text{ with rate
    }\frac{1}{1-X_{T-\beta}}, & \quad(2)
\end{aligned}
\end{equation}
respectively. For transitions in the process $\xi^{\mathcal X}$ due to
recombination we focus on geometry (i) first. A recombination event
hits one line between the $S$ and the $L$ locus with rate
$\rho_{SL}$ and between the $L$ and the $R$ locus with rate
$\rho_{LR}$. If a recombination event occurs between the $S$ and the
$L$ locus, it may be that both recombining chromosomes carry the same
allele at the $S$ locus. This gives a recombination event which cannot
be seen effectively and we ignore it in the process $\xi^{\mathcal
  X}$. All other recombination events must be modeled. If
$\xi_\beta^{\mathcal X}=(\xi^B,\xi^b)$ with $\xi^B=\{\xi^B_1, \ldots,
\xi^B_{|\xi^B|}\}$ and $\xi^b=\{\xi^b_1, \ldots, \xi^b_{|\xi^b|}\}$,
transitions occur for $1\leq j\leq |\xi^B|$ and $1\leq k\leq |\xi^b|$
from $(\xi^B, \xi^b) $ to
\begin{equation}\label{eq:rec1}
\begin{aligned}
  &  \left(\xi^B\setminus\{\xi^B_j\}, \xi^b\cup \{\xi^B_j\}\right) && \text{ with rate }\rho_{SL}(1-X_{T-\beta}) & \qquad (3_i) \\
  &  \left((\xi^B\setminus\{\xi^B_j\}) \cup \{\xi^B_j\cap \dickm \ell\}, \xi^b\cup \{\xi^B_j\cap\dickm r\} \right) && \text{ with rate }\rho_{LR}(1-X_{T-\beta}) & (4_i)\\
  &  \left((\xi^B\setminus \{\xi^B_j\}) \cup \{\xi^B_j\cap \dickm \ell, \xi^B_j\cap \dickm r\}, \xi^b\}\right)  && \text{ with rate }\rho_{LR}X_{T-\beta} & (5_i)\\
  &  \left(\xi^B, (\xi^b \setminus\{\xi^b_k\}) \cup \{\xi^b_k\cap \dickm \ell, \xi^b_k\cap\dickm r\}\right) && \text{ with rate }\rho_{LR}(1-X_{T-\beta}) & (6_i)\\
  &  \left(\xi^B\cup \{\xi^b_k\}, \xi^b\setminus\{\xi^b_k\}\right) && \text{ with rate }\rho_{SL}X_{T-\beta} & (7_i)\\
  & \left(\xi^B\cup \{\xi^b_k\cap\dickm r\}, (\xi^b\setminus\{\xi^b_k\}) \cup
  \{\xi^b_k\cap \dickm \ell\}\right) && \text{ with rate
  }\rho_{LR}X_{T-\beta}. & (8_i)
\end{aligned}
\end{equation}
Here, $(3_i)$ encodes a recombination event which takes a pair of
linked $L$- and $R$-loci from the beneficial to the wild-type
background; an event ($4_i$) separates the $R$-locus of a line and
takes it to the wild-type background; by $(5_i)$ the $L$ and $R$ loci
of a line in the beneficial background are split but remain both in
the same background; $(6_i)$ describes the same transition for a line
in the wild-type background.  The transitions $(7_i)$ and $(8_i)$
describe the back-recombination of loci into the beneficial
background.

\begin{figure}
\begin{center}
\includegraphics[width=15.5cm]{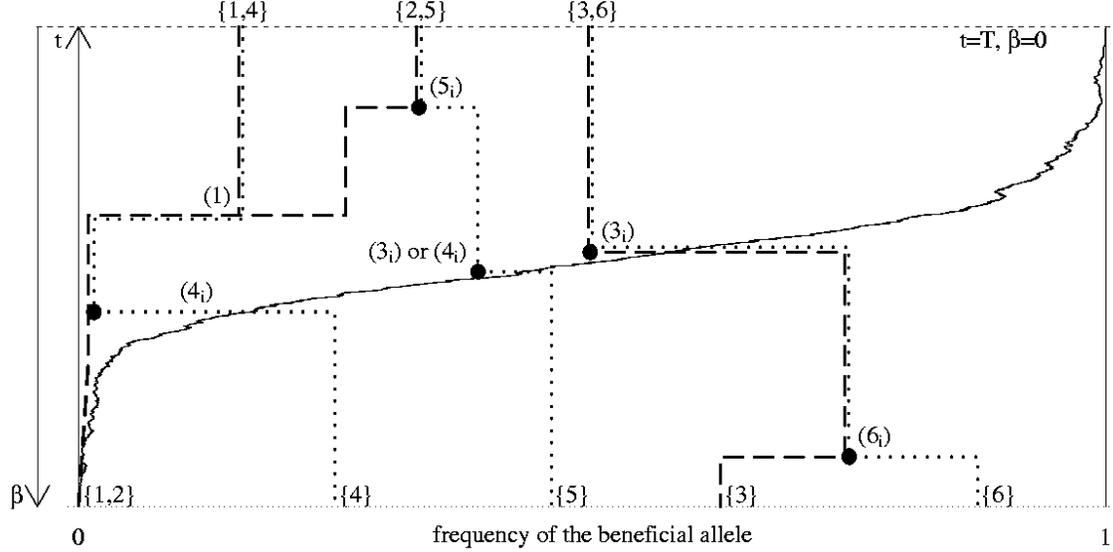}
\end{center}
\caption{\label{ancrecgraph}A structured ancestral recombination graph
  $\xi^{\mathcal X}$ conditioned on the frequency path $\mathcal X$ of
  the beneficial allele. Between times $\beta=0$ and $\beta=T$
  coalescences may occur at rates $(1)$ and $(2)$. Recombination
  events happen at rates $(3_i)-(8_i)$. The dashed lines indicate
  ancestry of the $L$-locus while the $R$-locus may be traced along
  dotted lines.}
\end{figure}

\begin{example}
  An example displaying the dynamics of the process $\xi^{\mathcal X}$
  for geometry (i) is shown in Figure \ref{ancrecgraph}.  The sets of
  $L$- and $R$-loci are $\dickm\ell = \{1,2,3\}$ and $\dickm
  r=\{4,5,6\}$, respectively. The starting partition is $\xi^{\mathcal
    X}_0=(\pi,\varnothing)$ with $\pi = \{\{1,4\}, \{2,5\},
  \{3,6\}\}$. Several kinds of events can happen; coalescences in the
  beneficial background, i.e., an event (1), recombinations which
  leave the two neutral loci together but change the allele at the
  selected site, i.e., an event $(3_i)$ and recombination events which
  split the two neutral loci. The last kind of event may either bring
  one of the two neutral loci in a different background, $(4_i)$, or
  split a line within the beneficial background, $(5_i)$, or split a
  line in the wild-type background, $(6_i)$. The final partition is
  $\xi^{\mathcal X}_T = (\xi^B_T, \xi^b_T)$ with $\xi^B_T =
  \{\{1,2\}\}$, $\xi^b_T= \{\{3\}, \{4\}, \{5\}, \{6\}\}$.
\end{example}

For geometry (ii) we have (rescaled) recombination rates $\rho_{LS}$
and $\rho_{SR}$ between the left neutral and the selective and the
right and the selective locus, respectively. Here, transitions occur
from $(\xi^B, \xi^b)$ to
\begin{equation}\label{eq:rec2}
\begin{aligned}
  &\left((\xi^B\setminus\{\xi^B_j\})\cup\{\xi_j^B\cap\dickm r\}, \xi^b\cup\{\xi_j^B\cap \dickm \ell\} \right) && \text{ with rate }\rho_{LS}(1-X_{T-\beta}) &  \quad (3_{ii})\\
  &\left((\xi^B\setminus\{\xi^B_j\})\cup\{\xi_j^B\cap\dickm \ell\}, \xi^b\cup\{\xi_j^B\cap \dickm r\} \right) && \text{ with rate }\rho_{SR}(1-X_{T-\beta}) & (4_{ii})\\
  &\left((\xi^B\setminus\{\xi^B_j\}) \cup \{\xi^B_j\cap \dickm \ell, \xi^B_j\cap \dickm r\}, \xi^b\right) && \text{ with rate }(\rho_{LS} + \rho_{SR})X_{T-\beta} & (5_{ii})\\
  &\left(\xi^B, (\xi^b\setminus\{\xi^b_k\}) \cup \{\xi^b_k\cap \dickm \ell, \xi^b_k\cap \dickm r\}\right) && \text{ with rate }(\rho_{LS} + \rho_{SR})(1-X_{T-\beta}) & (6_{ii})\\
  &\left(\xi^B\cup\{\xi_k^b\cap \dickm \ell\}, (\xi^b\setminus\{\xi^b_k\})\cup\{\xi_k^b\cap\dickm r\}\right)  && \text{ with rate }\rho_{LS}X_{T-\beta} & (7_{ii})\\
  &\left(\xi^B\cup\{\xi_k^b\cap \dickm r\},
 ( \xi^b\setminus\{\xi^b_k\})\cup\{\xi_j^b\cap\dickm \ell\}\right) && \text{
    with rate }\rho_{SR}X_{T-\beta}. & (8_{ii})
\end{aligned}
\end{equation}
These events refer to a change in background from the beneficial to
the wild-type background either for the $L$-locus, $(3_{ii})$, or the
$R$-locus, $(4_{ii})$. Splits in the beneficial and wild-type
background may happen as in the case of geometry (i); see events
$(5_{ii})$ and $(6_{ii})$. Back-recombinations to the beneficial
background are denoted by $(7_{ii})$ for the $L$- and $(8_{ii})$ for the
$R$-locus. Observe that a transition which takes both loci on one line
from the beneficial to the wild-type background cannot occur for
geometry (ii); cf. event $(3_i)$.

\begin{definition}\label{def:1}
  Assume $\dickm \ell$ and $\dickm r$ are sets of left and right
  neutral loci, respectively, and $\mathcal X=(X_t)_{0\leq t\leq T}$
  is a frequency path of the beneficial allele given by
  \eqref{eq:SDE}.

  Conditioned on $\mathcal X$, consider the jump process
  $\xi^{\mathcal X}=(\xi^{\mathcal X}_\beta)_{0\leq \beta \leq T}$,
  which starts in $\xi_0^{\mathcal X} = (\pi, \varnothing)$ for
  $\pi\in \mathcal P_{\dickm\ell \cup \dickm r}$ and makes transitions
  by coalescence events (1), (2), given by \eqref{eq:coal} and
  recombination events ($3_i$)-($8_i$) or ($3_{ii}$)-($8_{ii}$) from
  \eqref{eq:rec1} and \eqref{eq:rec2}, respectively. This process
  $\xi^{\mathcal X}$ is denoted the \emph{structured ancestral recombination graph for
    the $L$ and $R$ locus} conditioned on $\mathcal X$ for geometry
  (i) or (ii), respectively.

  The mixture of $\xi^{\mathcal X}_T$ over the distribution of
  frequency paths given by \eqref{eq:SDE} defines the random partition
  $\Gamma_\pi = (\Gamma^B_\pi, \Gamma^b_\pi)$, i.e.,
    $$\Gamma_\pi :=\int \xi^{\mathcal X}_T \mathbb{P}\left[d\mathcal X\right].$$
  \end{definition}

\section{Main result}
We study selective sweeps in the infinite population limit, i.e., the
frequency of the beneficial allele follows the SDE given by
\eqref{eq:SDE}. Moreover, selection is most efficient for large
selection coefficients. Our goal is to derive a simpler but approximate expression
for $\Gamma_\pi$ in the regime of large $\alpha$. It was shown in
\cite{EtheridgePfaffelhuberWakolbinger2006} that for the fixation time
$T$ of the beneficial allele
\begin{align}\label{eq:T}
  \mathbb E[T] = \frac{2\log\alpha}{\alpha} + \mathcal
  O\Big(\frac{1}{\alpha} \Big), \qquad \mathbb V[T] = \mathcal
  O\Big(\frac{1}{\alpha^2} \Big)
\end{align}
for large $\alpha$. This suggests that only under the scaling $\rho =
\mathcal O(\alpha/ \log\alpha)$ for the recombination rate a
non-trivial number of recombination events occurs during the sweep for large $\alpha$. This is true for all possible kinds of
recombination events during the sweep, so the recombination rates
$\rho_{SL}, \rho_{LR}$ and $\rho_{LS}, \rho_{SR}$ for geometries (i)
and (ii) should be of this order.  Henceforth, we assume
\begin{equation*}
  \begin{aligned}
    \text{Geometry (i):} &\qquad \rho_{SL} =
    \gamma_{SL}\frac{\alpha}{\log\alpha}, &\quad \rho_{LR} =
    \gamma_{LR}\frac{\alpha}{\log\alpha}, &\qquad 0<\gamma_{SL}, \gamma_{LR}<\infty\\
    \text{Geometry (ii):} &\qquad \rho_{LS} =
    \gamma_{LS}\frac{\alpha}{\log\alpha},&\quad \rho_{SR} =
    \gamma_{SR}\frac{\alpha}{\log\alpha},&\qquad 0<\gamma_{LS},
    \gamma_{SR}<\infty.
  \end{aligned}
\end{equation*}

Our approximation of $\Gamma_\pi$ is based on a Yule tree, which
serves as an approximation of the genealogy at the selected locus.  A
Yule tree is the realization of a Yule process, i.e., a pure birth
process which starts with one line and every line splits in two lines
after an exponential waiting time.

In our approximation the quantity
\begin{align}\label{eq:pjk}
  p_{i_1}^{i_2}(\gamma) := \exp\Big( -\frac{\gamma}{\log\alpha} \sum_{i=i_1+1}^{i_2}
  \frac{1}{i}\Big)
\end{align}
will play an important role. 

Assume $\dickm\ell$ and $\dickm r$ are sets of left and right loci and
$\pi\in\mathcal P_{\dickm \ell\cup\dickm r}$. Three mechanisms
determine the Yule approximation of the partition $\Gamma_\pi$. First,
we approximate splits in the beneficial background, i.e., events
$(5_i)$ and $(5_{ii})$, by the following procedure:
\begin{align}\label{eq:Y2}
  \text{\parbox{13cm}{For all partition elements
      $\pi_1,\ldots,\pi_{|\pi|}$ realize Bernoulli random variables
      $U_1,\ldots, U_{|\pi|}$ which are 1 with success probability
      $$ \text{geometry (i):} \quad 1-p_0^{\lfloor
        2\alpha\rfloor}(\gamma_{LR}))\qquad \text{geometry (ii):}\quad 
      1-p_0^{\lfloor 2\alpha\rfloor}(\gamma_{LS} + \gamma_{SR})).$$ If
      $U_i=1$, split the $i$th partition element in its left and right
      locus. Altogether, this defines a partition
$$ \pi' = \big\{ \{\pi_i \cap \dickm\ell\}, \{\pi_i\cap\dickm r\}: U_i=1\big\} \cup \big\{ \{\pi_i\}: U_i=0\big\}.$$
}}
\end{align}
Next, realize a Yule process with branching rate $\alpha$, i.e., each
line splits in two lines at rate $\alpha$. Stop this process when it
has $\lfloor {2}\alpha\rfloor$ lines. Call this tree $\mathcal Y$. To
obtain the genealogy of a sample of size $|\pi'|$ from this tree with
$\lfloor {2}\alpha\rfloor$ extant leaves, we use the following
construction:
\begin{align}\label{eq:Y1}
  \text{\parbox{13cm}{Start with $|\pi'|$ lines from the full Yule
      tree $\mathcal Y$ with $\lfloor 2\alpha \rfloor$ lines. When
      there are $k$ lines left at the time the full tree has $i$
      lines, the probability that a coalescence event occurs among the
      $k$ lines at the time the full tree goes from $i$ to $i-1$ lines
      is
      $$\frac{\binom{k}{2}}{\binom{i}{2}}.$$ By this construction we
      build a tree $\mathcal Y_{|\pi'|}$ with the partition elements
      of $\pi'$ as leaves and nodes which record the number of lines
      in the full Yule tree.  }}
\end{align}

\begin{remark}
  To construct the sample tree $\mathcal Y_{|\pi'|}$ from $\mathcal Y$
  is a task equivalent to describing an exchangeable sample from a
  tree which arises by exchangeable binary coalescence dynamics. This
  has been studied by \cite{Saundersetal1984} and was recalled in
  \cite[Lemma 4.8]{EtheridgePfaffelhuberWakolbinger2006}.  If $I_t=i$
  is the number of lines in the Yule tree $\mathcal Y$ at time $t$,
  denote by $K_i$ the number of lines in $\mathcal Y_{|\pi'|}$ while
  $I_t=i$. The process $(K_i)_{\lfloor 2\alpha \rfloor\geq i\geq 1}$
  is a time-inhomogeneous Markov chain with transition probabilities
$$ \mathbb P[K_{i-1}=k-1 | K_i=k] = \frac{\binom{k}{2}}{\binom{i}{2}},
\qquad \qquad i=2,\ldots, \lfloor 2\alpha\rfloor, k=2,\ldots, |\pi'|.
$$
Moreover, the sample tree can be described forward in time by noting
that
$$ \mathbb P[K_{i}=k|K_{i-1}=k-1 ] = \frac{|\pi'| - k+1}{|\pi'|+i-1}. $$
\qed
\end{remark}
 
The sample tree which is pruned out of the full tree in this way
represents the genealogy at the selected site. To describe the
genealogies at the partially linked neutral sites we mark the sample
Yule tree to determine further recombination events. A mark stands for
one (or two) recombination events that may occur. This works in the following way:
\begin{align}\label{eq:Y3}
  \text{\parbox{13cm}{Let a branch in the tree $\mathcal Y_{|\pi'|}$
      be given which starts when the full tree has $i_1$ lines and
      ends when the full genealogy has $i_2$ lines. For geometry (i),
      every branch can be hit by at most one of three different kinds
      of marks indicating recombination events.  These are $SL$-,
      $LR$-, and $SLR$-marks. Their probabilities are given in Table
      \ref{tab:marks}. For geometry (ii) the branch is hit
      independently by $LS$- and $SR$-marks with probabilities
      $(1-p_{i_1}^{i_2}(\gamma_{LS}))$ and $(1-p_{i_1}^{i_2}(\gamma_{SR}))$.\\[0.5ex]
      Here, $SL$-marks separate the $S$- from the $L$-locus on each
      branch of the tree, etc. For geometry (i), $SLR$-marks separate
      the $S$- from the $L$- and the $L$- from the $R$-locus.}}
\end{align}

\begin{table}
\begin{center}
\vspace{1ex}

\begin{tabular}{|c|c|}\hline
  \rule[-4mm]{0cm}{1cm}mark & probability \\\hline
  \rule[-4mm]{0cm}{1cm}$SL$ & $\big(1-p_{i_1}^{i_2}(\gamma_{SL})\big)p_0^{i_2}(\gamma_{LR})$ \\
  \rule[-4mm]{0cm}{1cm}$LR$ & $p_{i_1}^{i_2}(\gamma_{SL})\big(1-p_{i_1}^{i_2}(\gamma_{LR})\big)$ \\
  \rule[-4mm]{0cm}{1cm}$SLR$ & $\big(1-p_{i_1}^{i_2}(\gamma_{SL})\big)\big(1-p_0^{i_2}(\gamma_{LR})\big)$\\ 
  \rule[-4mm]{0cm}{1cm}no &  $p_{i_1}^{i_2}(\gamma_{SL})p_{i_1}^{i_2}(\gamma_{LR})$\\\hline
\end{tabular}
\end{center}
\caption{\label{tab:marks}For geometry (i), we mark every branch in
  the Yule tree by at most one from three different kinds of events.
  If a branch starts when the full Yule tree has $i_1$ and ends when
  it has $i_2$ lines, the probabilities for all marks are given in the
  table. }
\end{table}

\begin{figure}
\begin{center}
\includegraphics[width=15.5cm]{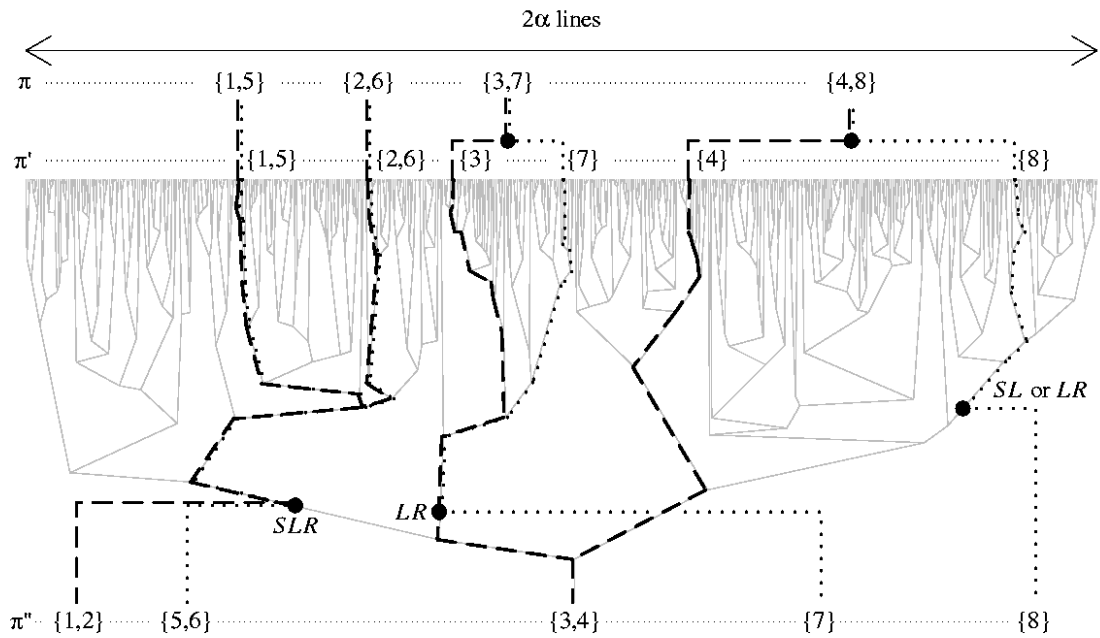}
\end{center}
\caption{\label{yuleTree}The Yule process approximation for two linked
  neutral loci under a selective sweep. Here, we consider geometry
  (i). The $L$-locus may be traced back along dashed lines while
  dotted lines indicate ancestry of the $R$-locus. See text for
  explanation. }
\end{figure}

\begin{example}
  The above construction is illustrated in Figure \ref{yuleTree}. We
  consider geometry (i) here. A set $\dickm\ell=\{1,2,3,4\}$ of
  $L$-loci and $\dickm r=\{5,6,7,8\}$ of $R$-loci is given. Starting with
  $\pi=\{\{1,5\}, \{2,6\},\{3,7\},\{4,8\}\}$, every partition element
  is split with probability $p_0^{\lfloor 2\alpha\rfloor}$ according
  to \eqref{eq:Y2}. This results in the finer partition $\pi'$. The
  partition elements of $\pi'$ are used to construct a sample tree
  from a full Yule tree which has $\lfloor 2\alpha\rfloor$ lines. The
  coalescence probabilities for the sample are given by \eqref{eq:Y1}.
  On the sample tree, branches are marked by $SL$-, $LR$-, or
  $SLR$-marks according to Table \ref{tab:marks}. The resulting
  partition $\pi''$ is constructed as given in Definition \ref{def:2}.
\end{example}

We are now in a position to define our approximation based on the Yule
process.

\begin{definition}\label{def:2}
  Assume $\dickm\ell$ and $\dickm r$ are sets of left and right
  neutral loci, respectively, and $\pi\in\mathcal P_{\dickm
    \ell\cup\dickm r}$. By \eqref{eq:Y2} construct the partition
  $\pi'$ and by \eqref{eq:Y1} and \eqref{eq:Y3} a Yule tree $\mathcal
  Y_{|\pi'|}$ with marks. For geometry (i) define the equivalence
  relation:
  \begin{equation}\label{eq:equivGeoi} 
    j\sim k :\iff \begin{cases} \text{no $SL$-, $SLR$-mark on } \Yup & \text{ if }j,k\in\dickm \ell,\\
      \text{no $SL$-, $LR$-, $SLR$-mark on } \Yup, & \text{ if }j,k\in\dickm r\\[2ex]
      \text{no $SL$-mark on } \Yup, \\[1ex]
      \text{no $LR$-mark on } \Yri, & \text{ if }j\in\dickm \ell,k\in\dickm r\\
      \text{no $SLR$-mark on } \Yall
  \end{cases}
  \end{equation}
  where the bold lines indicate for which part of the tree $\mathcal
  Y_{|\pi'|}$ relating two lines with the root of the tree, the
  constraint on marks applies. For geometry (ii) set
  \begin{equation}\label{eq:equivGeoii} 
    j\sim k :\iff \begin{cases} \text{no $LS$-mark on } \Yup, & \text{ if }j,k\in\dickm \ell,\\
      \text{no $SR$-mark on } \Yup, & \text{ if }j,k\in\dickm r,\\[2ex]
      \text{\parbox{5.2cm}{no $LS$-mark on \Yle, \\[1ex]
        no $SR$-mark on \Yri}} & \text{ if
      }j\in\dickm \ell, k\in\dickm r
    \end{cases} 
  \end{equation}
  (The equations \eqref{eq:equivGeoi} and \eqref{eq:equivGeoii} indeed
  define equivalence relations, as can easily be checked.)  Each of
  these equivalence relations on $\dickm\ell\cup\dickm r$ defines a
  partition $\pi''$.  For geometry (i) there is a unique partition
  element
\begin{equation}\label{eq:uni} 
\begin{aligned}
  \pi''_f = \Big\{j\in\dickm \ell: & \text{ no $SL$-, $SLR$-mark on
    \one{$\pi'_{(j)}$}}\Big\} \\ & \cup \Big\{k\in\dickm r: \text{ no $SL$-, $LR$-,
    $SLR$-mark on \one{$\pi'_{(k)}$}}\Big\}
\end{aligned}
\end{equation}
and for geometry (ii) a unique partition element
\begin{equation}
\begin{aligned}\label{eq:unii} 
  \pi''_f = \Big\{j\in\dickm \ell: & \text{ no $LS$-mark on
    \one{$\pi'_{(j)}$}}\Big\} \cup \Big\{k\in\dickm r: \text{ no $SR$-mark on
    \one{$\pi'_{(k)}$}}\Big\}.
\end{aligned}
\end{equation}
Then the random partition
$$ \Upsilon_\pi:= ( \{\pi''_f\}, \pi''\setminus \{\pi_f''\})$$
is called the \emph{Yule approximation of } $\Gamma_\pi$.
\end{definition}

\begin{example}
  For the example in Figure \ref{yuleTree} the $SL$-, $LR$- and
  $SLR$-marks on the sample tree lead to the realization
$$ \Upsilon_\pi = (\{\{3,4\}\}, \{\{1,2\}, \{5,6\},\{7\},\{8\}\}).$$
\end{example}

\begin{theorem}\label{T}
  Let $\pi\in\mathcal P_{\dickm \ell\cup\dickm r}$ and $\Gamma_\pi$
  and $\Upsilon_\pi$ be as in Definitions \ref{def:1} and \ref{def:2}.
  Then,
  $$ \sup_{\xi\in\mathcal P'_{\dickm \ell\cup\dickm r}} 
  \big|\mathbb P[\Gamma_\pi = \xi] - \mathbb P[\Upsilon_\pi=\xi] \big| =
  \mathcal O\Big( \frac{1}{(\log\alpha)^2}\Big). $$
\end{theorem}

\noindent

\begin{remark}
\begin{enumerate}
\item The Theorem states that, for large $\alpha$, the random
  partitions $\Gamma_\pi$ and $\Upsilon_\pi$ are close in variation
  distance. Here, variation distance refers to the maximal difference
  in the probabilities to obtain any partition $\xi\in\mathcal
  P'_{\dickm\ell \cup\dickm r}$. The order of accuracy, given by the
  Landau symbol, still depends on several parameters. These are the
  cardinalities $\ell$ and $r$ and recombination constants
  $\gamma_{SL}, \gamma_{LR}$ for geometry (i) and $\gamma_{LS}$ and
  $\gamma_{SR}$ for geometry (ii). The proof of Theorem \ref{T} will
  be given in Section \ref{proof}.
\item At first sight, comparing the Definitions \ref{def:2} and
  \ref{def:1} the Yule approximation does not look any simpler than
  the exact model. However, the Yule approximation has advantages both
  analytically and computationally. The random partition $\Gamma_\pi$
  relies on constructing a frequency path $\mathcal X$, while the Yule
  approximation $\Gamma_\pi$ constructs the ancestral recombination
  graph for the sample directly. Analytically, as we will see in
  Section \ref{app}, this means that explicit calculations are
  possible. Computationally, i.e., for simulations of the ancestral
  recombination graph, the direct construction of the ancestry of the
  sample allows for fast algorithms; see
  \cite{PfaffelhuberHauboldWakolbinger2006} for the case of a single
  neutral locus.
\item The current paper is a generalisation of results found in
  \cite{EtheridgePfaffelhuberWakolbinger2006} for a two-locus system
  with only one neutral locus. More precisely, consider the projection of
  $\Gamma_\pi$ on only one locus, i.e., on either $\dickm \ell$ or
  $\dickm r$. In Propositions 4.2 and 4.7 of that paper it was shown
  that the projection of $\Upsilon_\pi$ on $\dickm \ell$ or $\dickm r$
  is an approximation to a structured coalescent with an error in
  probability of the order $\mathcal O\big( (\log\alpha)^{-2}\big)$.
\item In \cite{EtheridgePfaffelhuberWakolbinger2006} an approximate
  sampling formula was given in the two-locus case. A similar approach
  would be possible here. However, we refrain from its derivation
  because it was shown in \cite{PfaffelhuberHauboldWakolbinger2006}
  that the sampling formula in the two-locus case only produces
  numerically sound results for $n\leq 5$.
\item As indicated numerically in
  \cite{PfaffelhuberHauboldWakolbinger2006}, the Yule approximation
  can be improved. To understand how this works, we need to collect
  the errors which contribute to the error of order $\mathcal
  O(1/(\log\alpha)^2)$. First, the Yule approximation ignores events
  $(2), (6_{ii}), (7)$ and $(8)$. Second, as will be clear in the
  proof of Proposition \ref{PropSecond}, the coalescent rate in the
  beneficial background is decreased from $1/X dt$ to $(1-X)/X dt$ by
  the Yule process. It is the latter error that dominates, at least in
  large samples, because the total coalescence rate increases
  quadratically with the number of lines. However, increasing the
  coalescence probability in \eqref{eq:Y1} to
\begin{align*}\label{eq:yuleCoal}
  1\wedge
  \frac{\binom{k}{2}}{\binom{i}{2}}\frac{1}{1-\tfrac{i-1}{2\alpha}}
\end{align*}
at the time the Yule tree has $i$ lines corrects for this error.
\item For simulations of genealogies it is most important that the
  Yule approximation given above is not restricted to the case of two
  neutral loci. The take-home-message from the construction of the
  Yule approximation is that splits in the beneficial background are
  generated first and afterwards marks on a Yule tree determine all
  recombination events. Both, splits in the beneficial background and
  recombination events along the Yule tree can be given along a
  continuous chromosome.
\end{enumerate}
\end{remark}
\qed

\section{Application: {\bf\emph{D}}}
\label{app}
Lewontin's $D$ is a measure of linkage disequilibrium (non-random
association of alleles) and is frequently used as a simple statistic
in a multi-locus setting (\cite{Lewontin1964}; see also
\cite[(2.89)]{Ewens2004}). Given two loci $L$ and $R$ with alleles 0
or 1 at each locus, it is defined as
\begin{equation} \label{eq:D}
D = p_{LR} - p_L p_R
\end{equation}
where $p_{LR}$ is the frequency of individuals carrying allele 1 at
both loci, $p_L$ is the frequency of 1's at the $L$ locus and $p_R$ is
the frequency of 1's at the $R$ locus..

To predict patterns of $D$ between pairs of neutral loci at the time
$T$ of fixation of a beneficial allele we next approximate $\mathbb
E[D(T)]$ using Theorem \ref{T}. It is crucial to observe that $\mathbb
E[p_{LR}(T)]$ as well as $\mathbb E[p_{L}(T) p_R(T)]$ may be derived
by the distribution of genealogies of linked neutral loci under
selection and the expected allele frequencies at the beginning of the
sweep. To see this, note that $\mathbb E[p_{LR}(T)]$ equals the
probability that the ancestors of the $L$- and $R$-locus of one
randomly picked individual from the population at time $T$ carry
alleles 1 at both neutral loci. Analogously, $\mathbb E[p_{L}(T)
p_R(T)]$ is the probability that the ancestors of the $L$- and $R$-
loci of two different individuals at time $T$ both carry allele 1.
Denote by $q$ the probability that both loci, $L$ and $R$ from one
individual, picked at time $T$, have a common ancestor at the
beginning of the sweep. Analogously, $q'$ is the same probability for
the $L$- and $R$-loci from two different individuals. Using these
definitions we see that
\begin{equation}\label{eq:pLRpLpR}
\begin{aligned}
  \mathbb E\left[p_{LR}(T)\right] & = q \cdot \mathbb E\left[p_{LR}(0)] + (1-q)\cdot
  \mathbb E[p_{L}(0)p_R(0)\right], \\ \mathbb E\left[p_{L}(T)p_R(T)\right] & = q'\cdot
  \mathbb E\left[p_{LR}(0)\right] + (1-q')\cdot \mathbb E\left[p_{L}(0)p_R(0)\right].
\end{aligned}
\end{equation}
Combining \eqref{eq:pLRpLpR} with the definition of $D$ from
\eqref{eq:D},
\begin{equation}\label{eq:D1}
\mathbb E[D(T)] = (q - q') \mathbb E[D(0)].
\end{equation}
Both, $q$ and $q'$ may be approximated by Theorem \ref{T}. Formally,
setting $\dickm{\ell}=\{1\}, \dickm{r}=\{2\}$,
\begin{equation}
\begin{aligned}
  q & = \mathbb P\left[\Gamma^B_{\{1,2\}} \cup \Gamma^b_{\{1,2\}}= \{\{1,2\}\}\right],\\
  q'& = \mathbb P\left[\Gamma^B_{\{1\},\{2\}} \cup \Gamma^b_{\{1\},\{2\}} =
  \{\{1,2\}\}\right].
\end{aligned}
\end{equation}
As $\Gamma_\pi$ may be approximated by $\Upsilon_\pi$ this brings us
in a position to predict patterns of $D$ at the end of a selective
sweep.

\begin{theorem}\label{T2}
For geometry (i), 
\begin{equation}\label{eq:P:D:1}
\begin{aligned}
  \mathbb E[D(T)] & = p_0^{ 2\alpha }(2\gamma_{LR}) \Big(1 -
  \sum_{k=2}^{ 2\alpha} \frac{2}{k(k+1)}
  p_k^{2\alpha}(2\gamma_{SL})\Big)\mathbb E[D(0)] + \mathcal O\Big(
  \frac{1}{(\log\alpha)^2}\Big),
\end{aligned}
\end{equation}
and for geometry (ii),
\begin{equation}\label{eq:P:D:2}
  \mathbb E[D(T)] = \mathbb E[D(0)]\cdot\mathcal O\Big(
  \frac{1}{(\log\alpha)^2}\Big).
\end{equation}
\end{theorem}

\begin{figure}
\begin{center}
\includegraphics[width=10cm]{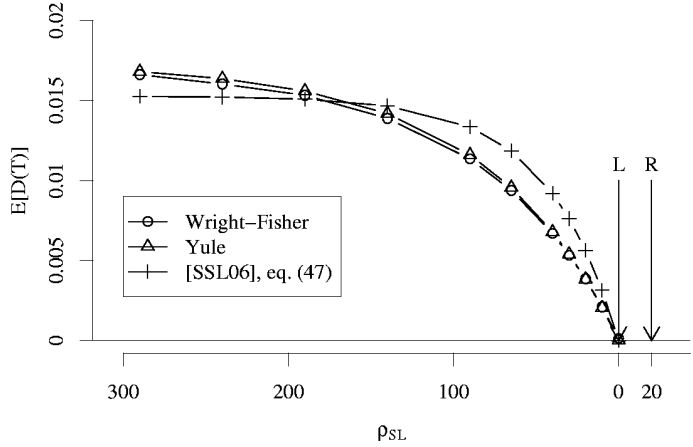}
\end{center}
\caption{\label{sim}The effect of Lewontin's $D$ under a selective
  sweep may be simulated in a Wright-Fisher model. In this process,
  the frequency path of the beneficial allele is stochastic and the
  ancestral recombination graph may be built conditioned on this
  frequency path. The locations of the $L$ and $R$ locus are fixed.
  The position of the selected site varies along the $x$-axis.  If we
  compare the result from \eqref{eq:P:D:1} to equation (47) of
  \cite{StephanSongLangley2006} we see that the Yule process
  approximation is more accurate. The parameters of the Wright-Fisher
  model are $N=10^5, \alpha=1000, \rho_{LR}=20$ and $D(0) = 0.0242$.}
\end{figure}

\begin{remark}
\begin{enumerate}
\item Patterns of Lewontin's $D$ can be studied by deterministic
  forward calculations instead of our genealogical approach. This was
  carried out in \cite{StephanSongLangley2006} under the assumption
  that strong selection leads to a deterministic behaviour of allele
  frequencies. Specifically, the frequency of the beneficial allele
  follows the logistic differential equation
  $$ dX = \alpha X(1-X)dt,\qquad\qquad X_0 = \tfrac{1}{N} $$
  instead of the stochastic path given by \eqref{eq:SDE}. Predictions
  of $D$ at all times during the selective sweep were given. In
  particular, their equation (47) approximates values of $D$ at the
  end of the sweep for geometry (i).

  In real populations, random effects due to genetic drift are not
  negligible.  This has been pointed out by
  \cite{LehnertStephanPfaffelhuber2006}.
  The Yule process approximation captures most random effects. Indeed,
  comparison with simulations from
  \cite{LehnertStephanPfaffelhuber2006} shows that the results
  produced by the Yule process approximation are more accurate than
  those of \cite{StephanSongLangley2006}.
 

\item For empirical studies it is most interesting to know which
  patterns of linkage disequilibrium  to look for in real data. The
  pattern genetic hitchhiking can produce was discussed in
  \cite{StephanSongLangley2006} and \cite{ReedTishkoff2006}.
  Surprisingly, hitchhiking reduces levels of linkage disequilibrium
  compared to the neutral expectation. This is evident from Figure
  \ref{sim}. If the selected locus is far from both neutral loci,
  linkage disequilibrium between the neutral loci is not affected by
  hitchhiking. Therefore, values of $D$ for large $\rho_{SL}$ converge
  to the expectation of $D$ under neutrality.  This effect was taken
  up by \cite{ReedTishkoff2006} to argue that genetic hitchhiking
  produces patterns in the association of alleles similar to
  recombination hotspots, which are e.g. important in genetic
  association studies in humans (\cite{hapmap2005}). However, genetic
  hitchhiking certainly produces patterns different from recombination
  hotspots in general, e.g., a low neutral diversity or a
  distinctive site frequency spectrum (\cite{FayWu2000}).
\item An accurate approximation of $\mathbb E[D(T)]$ does not suffice
  to predict patterns of linkage disequilibrium in general. In
  addition to genetic drift, random effects which affect $D(T)$ were
  found in \cite{StephanSongLangley2006} to be the allelic type of the
  founder of the sweep and its frequency. The resulting variance in
  $D$ can be considerably higher than under neutrality.
\end{enumerate}
\end{remark}

\noindent
Now we come to the proof of Theorem \ref{T2}.

\begin{proof} The key in the proof is to compute the probabilities $q$
  and $q'$. This is achieved by the Yule process approximation
  $\Upsilon_\pi$ of Theorem \ref{T}.

   We start with geometry (ii). Here, we can see from the Yule
   approximation \eqref{eq:equivGeoii} that $q = q'$ up to a term of
   order $1/(\log\alpha)^2$ since one $L$ and one $R$ locus are
   identical by descent iff there is no $LS$ mark on
   \text{\parbox{1.2cm}{
\beginpicture
\setcoordinatesystem units <0.1cm, 0.1cm>
\setplotarea x from 3 to 17, y from 3 to 17
\plot 10 6 10 10 7 13 10 10 13 13 /
\multiput{\tiny $\bullet$} at 10 10 *100  -0.03 0.03 /
\multiput{\tiny $\bullet$} at 10 10 *100  0 -0.04 /
\put{$\bullet$}[Cc]  at 10 6
\put{\footnotesize$1$} [cC] at 5 14
\put{\footnotesize$2$} [cC] at 15 14
\endpicture}} 
and no $SR$ mark on 
   \text{\parbox{1.2cm}{
\beginpicture
\setcoordinatesystem units <0.1cm, 0.1cm>
\setplotarea x from 3 to 17, y from 3 to 17
\plot 10 6 10 10 7 13 10 10 13 13 /
\multiput{\tiny $\bullet$} at 10 10 *100  0.03 0.03 /
\multiput{\tiny $\bullet$} at 10 10 *100  0 -0.04 /
\put{$\bullet$}[Cc]  at 10 6
\put{\footnotesize$1$} [cC] at 5 14
\put{\footnotesize$2$} [cC] at 15 14
\endpicture}}. It
   does not depend on the linkage of the $L$ and the $R$ locus at the
   end of the sweep.  Consequently, \eqref{eq:P:D:2} follows.

   For geometry (i), we start with the approximation of $q'$. For one
   $L$ and one $R$ locus from two different individuals there is a
   random number $K$ of lines in the full tree of the Yule
   approximation at the time the selected loci which are linked to the
   neutral ones coalesce.  To obtain the distribution of $K$, we
   compute
$$ \mathbb P[K=k] = \prod_{l=k+1}^{ 2\alpha} \left( 1 -
\frac{1}{\binom{l}{2}}\right) \frac{1}{\binom{k}{2}} = \left(
\prod_{l=k+1}^{ 2\alpha} \frac{(l+1)(l-2)}{l(l-1)}\right)
\frac{2}{k(k-1)} = \frac{2}{k(k+1)} + \mathcal O\left(
\frac{1}{\alpha}\right),$$ which is a special case of
\cite{EtheridgePfaffelhuberWakolbinger2006}, (4.16). We read from
\eqref{eq:equivGeoi} that the $L$ and $R$ locus are identical by
descent at the beginning of the sweep if and only if (a) no mark or an
$SL$ mark falls on 
   \text{\parbox{1.2cm}{
\beginpicture
\setcoordinatesystem units <0.1cm, 0.1cm>
\setplotarea x from 3 to 17, y from 3 to 17
\plot 10 6 10 10 7 13 10 10 13 13 /
\multiput{\tiny $\bullet$} at 10 10 *100  0 -0.04 /
\put{$\bullet$}[Cc]  at 10 6
\put{\footnotesize$1$} [cC] at 5 14
\put{\footnotesize$2$} [cC] at 15 14
\endpicture}}, 
(b) no mark hits
   \text{\parbox{1.2cm}{
\beginpicture
\setcoordinatesystem units <0.1cm, 0.1cm>
\setplotarea x from 3 to 17, y from 3 to 17
\plot 10 6 10 10 7 13 10 10 13 13 /
\multiput{\tiny $\bullet$} at 10 10 *100  0.03 0.03 /
\put{$\bullet$}[Cc]  at 10 6
\put{\footnotesize$1$} [cC] at 5 14
\put{\footnotesize$2$} [cC] at 15 14
\endpicture}} 
and (c) no mark or an $LR$ mark falls on
   \text{\parbox{1.2cm}{
\beginpicture
\setcoordinatesystem units <0.1cm, 0.1cm>
\setplotarea x from 3 to 17, y from 3 to 17
\plot 10 6 10 10 7 13 10 10 13 13 /
\multiput{\tiny $\bullet$} at 10 10 *100  -0.03 0.03 /
\put{$\bullet$}[Cc]  at 10 6
\put{\footnotesize$1$} [cC] at 5 14
\put{\footnotesize$2$} [cC] at 15 14
\endpicture}}. Hence we compute
\begin{equation}
\begin{aligned}
  q' & = \sum_{k=2}^{ 2\alpha} \frac{2}{k(k+1)}
  p_0^k(\gamma_{LR}) p_k^{ 2\alpha }(\gamma_{SL})p_k^{ 2\alpha
  }(\gamma_{LR})p_k^{ 2\alpha }(\gamma_{SL}) + \mathcal O\left(
  \frac{1}{(\log\alpha)^2}\right)\\ & = p_0^{ 2\alpha }(\gamma_{LR})
  \sum_{k=2}^{ 2\alpha} \frac{2}{k(k+1)} p_k^{2\alpha}(2\gamma_{SL}) +
  \mathcal O\left( \frac{1}{(\log\alpha)^2}\right).
\end{aligned}
\end{equation}
For $q$ we have to distinguish the cases where the $L$- and the $R$-loci
 split or not. If they do not split, the $L$- and $R$-locus have
the same ancestor at the beginning of the sweep if and only if there
is neither an $LR$- nor an $SLR$-mark on \one{$\{1,2\}$}. If they split,
the probability of a common ancestor is $q'$. Therefore,
\begin{equation}
\begin{aligned}
q & = p_0^{ 2\alpha }(\gamma_{LR}) p_0^{ 2\alpha }(\gamma_{LR}) +
\big(1-p_0^{ 2\alpha }(\gamma_{LR})\big) q'+
  \mathcal O\left( \frac{1}{(\log\alpha)^2}\right).
\end{aligned}
\end{equation}
Hence 
\begin{equation}
\begin{aligned}
  \mathbb E[D(T)] & = p_0^{ 2\alpha }(\gamma_{LR}) \big(p_0^{2\alpha
  }(\gamma_{LR}) - q'\big) \mathbb E[D(0)]+ \mathcal O\left(
  \frac{1}{(\log\alpha)^2}\right)
\end{aligned}
\end{equation}
and the result follows. 
\end{proof}

\section{Proof of Theorem  \ref{T}}
\label{proof}
The proof deals with geometries (i) and (ii) simultaneously.  We will
write events at rates (3)-(8) whenever we refer to the rates
($3_i$)-($8_i$) for geometry (i) and ($3_{ii}$)-($8_{ii}$) for
geometry (ii), respectively.

We will be dealing with several random partitions all of which agree
up to an error of order $\mathcal O\big( (\log(\alpha))^{-2}\big)$.
Exactly, we will prove
$$\Gamma_\pi \quad \stackrel{\text{Prop. \ref{PropFirst}}}{\approx}\quad 
\Delta_\pi \quad \stackrel{\text{Prop.
    \ref{PropSecond}}}{\approx}\quad \Xi_\pi \quad
\stackrel{\text{Prop. \ref{PropThird}}}{\approx}\quad \Upsilon_\pi$$ where
$\Gamma_\pi, \Delta_\pi, \Xi_\pi$ and $\Upsilon_\pi$ are given in
Definitions \ref{def:1}, \ref{def:3}, \ref{def:4} and \ref{def:2},
respectively and '$\approx$' means that the random partitions differ
by $\mathcal O\left( (\log\alpha)^{-2}\right)$ in variation distance.

While $\Gamma_\pi$ is the random partition which is defined by the
structured ancestral recombination graph, the other random partitions
are approximations. First, $\Delta_\pi$ arises by (i) ignoring events
which occur according to rates $(2), (6_{ii}), (7)$ and $(8)$ and (ii)
realizing all events according to rate $(5)$ first and only
afterwards, construct the process using rates $(1), (3), (4)$ and
$(6_i)$. Second, $\Xi_\pi$ already deals with the Yule process. It is
derived by marking an infinite Yule tree by two constant rate Poisson
processes with rates $\rho_{SL}, \rho_{LR}$ for geometry (i) and
$\rho_{LS}, \rho_{SR}$ for geometry (ii). Finally, the Yule
approximation $\Upsilon_\pi$ of $\Gamma_\pi$ arises by considering
only the number of lines in an infinite Yule tree at times of
coalescence in a sample.

In the whole proof we rely on a probability measure $\mathbb P$ on a
probability space on which the solution of \eqref{eq:SDE} as well as
arbitrarily many independent Poisson processes and other random
variables are realized.

\begin{definition}
\label{def:3}
Define a $\mathcal P'_{\dickm\ell\,\cup\,\dickm r}$-valued random
variable $\Delta_\pi$ as follows: starting in $\pi\in\mathcal
P_{\dickm \ell\cup\dickm r}$ split all partition elements $\xi\in\pi$
independently into $\xi \cap \dickm \ell, \xi \cap \dickm r$ with
probability
\begin{equation} \label{Split} 1-\mathbb E\left[\exp\left( - \rho\cdot
    \mathbb \int_0^T X_sds\right)\right]
\end{equation}
where $\rho= \rho_{LR}$ for geometry (i) and $\rho= \rho_{LS}+
\rho_{SR}$ for geometry (ii).  The resulting partition $\pi'$ is used
for the starting point $(\pi',\varnothing)$ of a process
$\eta^{\mathcal X} = (\eta^{\mathcal X}_\beta)_{0\leq\beta\leq T}$,
conditioned on a frequency path $\mathcal X = (X_t)_{0\leq t\leq T}$
with transitions according to events ($1$),($3_i$), ($4_i$), ($6_i$),
given by \eqref{eq:rec1}, for geometry (i) and to events ($1$),
($3_{ii}$) and ($4_{ii}$), given by \eqref{eq:rec2}, for geometry
(ii), respectively. Given $\eta^{\mathcal X}$, define $$\Delta_\pi :=
\int \eta^{\mathcal X}_T \mathbb P[d\mathcal X].$$
\end{definition}

\begin{proposition}\label{PropFirst}
  Let $\pi\in\mathcal P_{\dickm \ell\cup\dickm r}$ and $\Gamma_\pi$
  and $\Delta_\pi$ be as in Definitions \ref{def:1} and \ref{def:3}.
  Then,
  $$ \sup_{\xi\in\mathcal P'_{\dickm \ell\cup\dickm r}}
  \big|\mathbb P[\Gamma_\pi = \xi] - \mathbb P[\Delta_\pi=\xi] \big|
  = \mathcal O\left( \frac{1}{(\log\alpha)^2}\right). $$
\end{proposition}

\begin{proof}
  We proceed in several steps. Our arguments in Step 1 show that we
  may discard events which occur at rates (2), ($6_{ii}$), (7) and
  (8). In Step 2 we use a fixed number of Poisson processes to
  generate the random partition we want to approximate.  Our goal is
  to separate events ($5$) from the rest by verifying a certain order
  of the possible events and establishing an approximate independence
  of the events (5). Particularly, we show in Step 3 that splits in
  the beneficial background (i.e., events (5)) take place before all
  other events with high probability. The approximate independence
  will be proved in Steps 5 and 6 by an application of a general
  result on mixed Poisson processes we establish in Step 4.

  \begin{step}
    (Small probability of events (2), ($6_{ii}$), (7) and (8))\\
    First, note that by Proposition 3.4 of
    \cite{EtheridgePfaffelhuberWakolbinger2006} events ($2$), i.e.,
    coalescences in the wild-type background, have a probability of
    order $\mathcal O\big((\log\alpha)^{-2}\big)$.  Furthermore,
    events ($7$) and ($8$) are back-recombinations into the beneficial
    background and hence have a probability of order $\mathcal
    O\big((\log\alpha)^{-2}\big)$ as well.  Additionally, for geometry
    (ii), events ($6_{ii}$), i.e., splits in the wild-type background,
    can only occur if a coalescence event (2) has happened before.
    As a consequence, we can discard events which occur at rates (2),
    ($6_{ii}$), (7) and (8) producing only an error in variation
    distance of at most $\mathcal O\big((\log\alpha)^{-2}\big)$.

    So we are left with a $\mathcal P'_{\dickm\ell\cup\dickm
      r}$-valued stochastic process conditioned on $\mathcal X$,
    $\zeta^{\mathcal X} = (\zeta^{\mathcal X}_\beta)_{0\leq \beta\leq
      T}$, which arises by events $(1)$, $(3)$,$(4)$,$(5)$ and
    $(6_i)$, started in $\zeta_0^{\mathcal X} = (\pi,\varnothing)$.
  \end{step}

  \begin{step} (Construction of $\zeta^{\mathcal X}$ by Poisson processes)\\
    Recall that $\ell:=|\dickm {\ell}|$ and $r:=|\dickm{r}|$ are the
    number of $L$ and $R$ loci under consideration. Take Poisson
    processes which are all conditionally independent given the random
    frequency path $\mathcal X$ of the beneficial allele.  For
    coalescence, take a Poisson process $\mathcal T_{\mathfrak 1}$ with
  \begin{equation}
    \begin{aligned}
      \text{ rate }\binom{\ell + r}{2} \frac{1}{X_{T-\beta}} && 
      && \qquad
      (\text{coalescence in the beneficial background}) && \qquad \mathfrak{(1)},
    \end{aligned}
  \end{equation}
  at time $\beta$; for recombination events take Poisson processes
  $\mathcal T_{\mathfrak {3_i}}$, $\mathcal T_{\mathfrak {4_i}}$,
  $\mathcal T_{\mathfrak {5_i}}$ with
  \begin{equation}
    \begin{aligned}
      &\text{ rate }\ell\rho_{SL}(1-X_{T-\beta})& &\qquad(\text{rec. to the wild-type background})&& \qquad \mathfrak{(3_i)},\\
      & \text{ rate }  r \rho_{LR}(1-X_{T-\beta})& &\qquad(\text{rec. to or split in the wild-type background})&&\qquad \mathfrak{(4_i)},\\
      &\text{ rate } r \rho_{LR}X_{T-\beta} & &\qquad(\text{split in the beneficial background})&&\qquad \mathfrak{(5_i)},\\
    \end{aligned}
  \end{equation}
  at time $\beta$ for geometry (i) and Poisson processes $\mathcal
  T_{\mathfrak {3_{ii}}}$, $\mathcal T_{\mathfrak {4_{ii}}}$, $\mathcal
  T_{\mathfrak {5_{ii}}}$ with
  \begin{equation}
    \begin{aligned}
      &\text{ rate } \ell \rho_{LS}(1-X_{T-\beta})& &\qquad(\text{rec. to the wild-type background})&&\qquad \qquad \quad\: \mathfrak{(3_{ii})},\\
      & \text{ rate } r \rho_{SR}(1-X_{T-\beta})& &\qquad(\text{rec. to the wild-type background})&& \qquad \qquad \quad\: \mathfrak{(4_{ii})},\\
      &\text{ rate } r (\rho_{LS}+\rho_{SR})X_{T-\beta}&
      &\qquad(\text{split in the beneficial background})&& \quad
      \qquad \qquad \:\mathfrak{(5_{ii})},
    \end{aligned}
  \end{equation}
  at time $\beta$ for geometry (ii). We have combined recombinations
  to the wild-type and splits in the wild-type background in case of
  geometry $(i)$ since they happen with the same rates.

  Additionally, let $W=(W_{{\mathfrak i},m})_{{\mathfrak i} =
    {\mathfrak 1}, {\mathfrak 3}, {\mathfrak 4}, {\mathfrak 5},
    m=1,2,\ldots}$ be a random array such that all $W_{{\mathfrak
      i},m}$'s are independent, $W_{{\mathfrak 1},m}$ is uniformly
  distributed on all pairs of $\dickm{\ell}\cup\dickm{r}$,
  $W_{{\mathfrak 3},m}$ is uniformly distributed on $\dickm{\ell}$,
  and $W_{{\mathfrak 4},m}$ and $W_{{\mathfrak 5},m}$ are uniformly
  distributed on $\dickm{r}$, $m=1,2,\ldots$. 

  The set $\dickm{\ell}\cup\dickm{r}$ can be totally ordered, so we
  may assume that every partition element in $\zeta\in\mathcal
  P'_{\dickm{\ell}\cup \dickm{r}}$ has a smallest element. Recall that
  we write $\zeta_{(j)}$ for the partition element containing
  $j\in\dickm{\ell}\cup\dickm{r}$.

  We abbreviate by $\mathcal T_{\mathfrak 3}$-$\mathcal T_{\mathfrak
    5}$ the Poisson processes $\mathcal T_{\mathfrak {3_i}}$-$\mathcal
  T_{\mathfrak {5_i}}$ for geometry (i) and the Poisson processes
  $\mathcal T_{\mathfrak {3_{ii}}}$-$\mathcal T_{\mathfrak {5_{ii}}}$ for
  geometry (ii). We next show that the distribution of
  $\zeta^{\mathcal X}_T$ is the image measure of the tupel $(\mathcal
  T_{\mathfrak 1}, \mathcal T_{\mathfrak 3}, \mathcal T_{\mathfrak 4},
  \mathcal T_{\mathfrak 5}, W)$ under a map $\varphi$. Specifically,
  the distribution of $\zeta_T^{\mathcal X}$ is uniquely determined by
  the distribution of $(\mathcal T_{\mathfrak 1}, \mathcal
  T_{\mathfrak 3}, \mathcal T_{\mathfrak 4}, \mathcal T_{\mathfrak 5},
  W)$.

  To define $\varphi$, consider a discrete set $\mathbf T_{\mathfrak
    1}\subseteq[0,T]$ and finite sets $\mathbf T_{\mathfrak 3},\mathbf
  T_{\mathfrak 4},\mathbf T_{\mathfrak 5}\subseteq [0,T]$ such that
  $\mathbf T_{\mathfrak i_1}\cap \mathbf T_{\mathfrak i_2}=\varnothing$
  for $\mathfrak{i_1} \neq \mathfrak{i_2}$ and set $\mathbf
  T=\bigcup_{\mathfrak i} \mathbf T_{\mathfrak i}$. Furthermore $w =
  (w_{{\mathfrak i},m})_{{\mathfrak i} = {\mathfrak 1}, {\mathfrak 3},
    {\mathfrak 4}, {\mathfrak 5}, m=1,2,\ldots}$ such that for all
  $m=1,2,\ldots$, $w_{{\mathfrak 1},m}$ is a pair in
  $\dickm{\ell}\cup\dickm{r}$, $w_{{\mathfrak 3},m} \in \dickm{\ell}$
  and $w_{{\mathfrak 4},m}, w_{{\mathfrak 5},m} \in\dickm{r}$. Given
  $(\mathbf T_{\mathfrak 1},\mathbf T_{\mathfrak 3},\mathbf
  T_{\mathfrak 4},\mathbf T_{\mathfrak 5},w)$ we generate a partition
  by considering the events in $\mathbf T$ in decreasing order. Assume
  $\zeta^{\mathcal X}_0=(\pi,\varnothing)$ and after the $(m-1)$st
  event at time $\beta$ we obtain a partition $\zeta^{\mathcal
    X}_\beta = (\zeta^B, \zeta^b)\in\mathcal
  P'_{\dickm{\ell}\cup\dickm{r}}$ and the $m$th event in $\mathbf T$
  to be realized happens at time $\beta'\in \mathbf T$.

  Consider first the case $\beta'$ is the $m$th event is the
  $m_{\mathfrak 1}$st event in $\beta'\in \mathbf T_{\mathfrak 1}$.
  The pair $w_{{\mathfrak 1}, m_{\mathfrak 1}}=(j,k)$ gives a random
  pair of loci.  If $\zeta_{(j)}, \zeta_{(k)}\in\zeta^B$ and if both,
  $j$ and $k$, are the smallest elements of their partition elements,
  coalesce these partition elements, i.e., make the transition
  $$ \left(\zeta^B,\zeta^b\right) \longrightarrow \left((\zeta^B\setminus \{\zeta_{(j)}, 
  \zeta_{(k)}\}) \cup \{ \zeta_{(j)}\cup \zeta_{(k)}\},\; \zeta^b\right).$$
  Otherwise do nothing.

  The next case to consider is that $\beta'$ is the $m_{\mathfrak
    3}$rd event in $\mathbf T_{\mathfrak 3}$ and $w_{{\mathfrak 3},
    m_{\mathfrak 3}}=j$ for some $j\in\dickm{\ell}$. If
  $\zeta_{(j)}\in\zeta^B$ and if $j$ is the smallest element of
  $\zeta_{(j)}\cap\dickm{\ell}$, change the partition element from
  $\zeta^B$ to $\zeta^b$, i.e., make the transition
  \begin{equation}\label{eq:trans3}
    \left(\zeta^B,\zeta^b\right) \longrightarrow \left(\zeta^B\setminus \{\zeta_{(j)}\},\; \zeta^b\cup\{\zeta_{(j)} \}\right). 
  \end{equation}
  Otherwise do nothing. The case $t\in \mathbf T_{\mathfrak 5}$ is similar and
  is omitted.

  If $\beta'$ is the $m_{\mathfrak 4}$th event in $\mathbf
  T_{\mathfrak 4}$ and $w_{{\mathfrak 4}, m_{\mathfrak 4}}=j$ for
  $j\in\dickm{r}$ the partition $\zeta$ again only changes if $j =
  \min \zeta_{(j)}\cap\dickm{r}$. We distinguish two cases,
  $\zeta_{(j)}\in\zeta^B$ and $\zeta_{(j)}\in\zeta^b$. In the former
  case, split the $L$- and $R$-loci in the partition element in two
  partition elements and bring all $R$-loci into the wild-type
  background, i.e., make the transition
  \begin{equation}\label{eq:trans4a} \left(\zeta^B,\zeta^b\right) \longrightarrow 
    \left((\zeta^B\setminus\{\zeta^B_{(j)}\}) \cup \{\zeta^B_{(j)}\cap \dickm
    \ell\},\; \zeta^b\cup \{\zeta^B_{(j)}\cap\dickm r\}\right).\end{equation} 
  This corresponds to an event (4).   
  In the latter case split all $L$- and $R$-loci of $\zeta_{(j)}$ and leave them in
  the wild-type background, i.e., make the transition
  \begin{equation}\label{eq:trans4b} 
  \left(\zeta^B,\zeta^b\right) \longrightarrow \left(\zeta^B,Ê\;
    (\zeta^b\setminus\{\zeta_{(j)}\}) \cup \{\zeta_{(j)} \cap
    \dickm{\ell}, \zeta_{(j)}\cap\dickm{r}\}\right),
    \end{equation}
  which corresponds to an event $(6_i)$. 
  Recall that for geometry (ii) one $L$- and one $R$-locus cannot recombine to the wild-type background
  together.  Hence partition elements in $\zeta^b$ are either subsets of 
  $\dickm{\ell}$ or of $\dickm{r}$ such that the last transition must not occur for this geometry.

  By generating all events according to this procedure we end with a
  partition $\zeta^{\mathcal X}_T$. Therefore we have defined the map
  $\varphi: (\mathbf T_{\mathfrak 1},\mathbf T_{\mathfrak 3},\mathbf
  T_{\mathfrak 4},\mathbf T_{\mathfrak 5},w) \mapsto \zeta^{\mathcal
    X}_T$.
  \begin{align}\label{eq:claim}
    \text{\parbox{12cm}{\it The distribution of $\zeta^{\mathcal X}_T$ is
        the image measure of $(\mathcal T_{\mathfrak 1}, \mathcal
        T_{\mathfrak 3}, \mathcal T_{\mathfrak 4}, \mathcal
        T_{\mathfrak 5}, W)$ under the map $\varphi$.}}
  \end{align}
  To see this, observe first, that there are only finitely many
  recombination events (3), (4), (5) and ($6_i$). Almost surely, all
  events in the Poisson processes occur at different times, so
  $\varphi$ is defined on a set of probability 1.  By the above
  construction, we obtain that two partition elements in $\zeta^B$
  coalesce by event (1). The Poisson processes $\mathcal T_{\mathfrak
    1}$, $\mathcal T_{\mathfrak 3}$, $\mathcal T_{\mathfrak 4}$,
  $\mathcal T_{\mathfrak 5}$ produce exactly the recombination events
  (3), (4), (5) and ($6_i$). Hence \eqref{eq:claim} is proved.
  \smallskip
  
  Given $w$, the random partition $\varphi(\mathbf T_{\mathfrak 1},
  \mathbf T_{\mathfrak 3}, \mathbf T_{\mathfrak 4}, \mathbf
  T_{\mathfrak 5}, w)$ only depends on the order of time points in
  $\mathbf T_{\mathfrak 1}, \mathbf T_{\mathfrak 3}, \mathbf
  T_{\mathfrak 4}, \mathbf T_{\mathfrak 5}$. There is another feature
  we will need:
  \begin{align}\label{eq:claim2}
    \text{\parbox{12cm}{\it Let $\beta',\beta''$ be consecutive time
        points in $\mathbf T$ with $\beta'\in \mathbf T_{\mathfrak 3},
        \beta''\in \mathbf T_{\mathfrak 4}$.  Exchanging $\beta'$ and
        $\beta''$ does not alter the random partition $\varphi(\mathbf
        T_{\mathfrak 1}, \mathbf T_{\mathfrak 3}, \mathbf T_{\mathfrak
          4}, \mathbf T_{\mathfrak 5}, w)$. Formally, if $\mathbf T
        \cap (\beta',\beta'')=\varnothing$, $\mathbf T_{\mathfrak 3}' =
        \mathbf T_{\mathfrak 3}\setminus \{\beta'\} \cup \{\beta''\}$
        and $\mathbf T_{\mathfrak 4}' = \mathbf T_{\mathfrak
          4}\setminus \{\beta''\} \cup \{\beta'\}$. Then $$
        \varphi(\mathbf T_{\mathfrak 1}, \mathbf T_{\mathfrak 3}',
        \mathbf T_{\mathfrak 4}', \mathbf T_{\mathfrak 5}, w) =
        \varphi(\mathbf T_{\mathfrak 1}, \mathbf T_{\mathfrak 3},
        \mathbf T_{\mathfrak 4}, \mathbf T_{\mathfrak 5}, w). $$}}
  \end{align}
  Assume $\beta'$ is the $m_{\mathfrak 3}$rd event in $\mathbf
  T_{\mathfrak 3}$, $w_{{\mathfrak 3},m_{\mathfrak 3}}=j$ and
  $\beta''$ is the $m_{\mathfrak 4}$th event in $\mathbf T_{\mathfrak
    4}$ and $w_{{\mathfrak 4},m_{\mathfrak 4}}=m$. If $j$ and $k$ are
  not in the same partition element for $\beta<\beta'$, the claim is
  trivial as recombination events only make the partition finer.
  Similarly, if $j>\min \zeta_{(j)}\cap\dickm{\ell}$ or $k>\min
  \zeta_{(k)}\cap\dickm{r}$ only one transition occurs and the claim
  follows. In the case $$\zeta_{(j)} = \zeta_{(k)},\quad j = \min
  \zeta_{(j)}\cap\dickm{\ell},\quad k = \min
  \zeta_{(j)}\cap\dickm{r}$$ two transitions occur if and only if
  $\zeta_{(j)}=\zeta_{(k)} \in \zeta^B$. We illustrate this situation
  in Figure \ref{smallFig}.

  \begin{figure}
    \hspace{3cm} (a) \hspace{7.5cm}(b)

  \begin{center}
    \includegraphics[width=7cm]{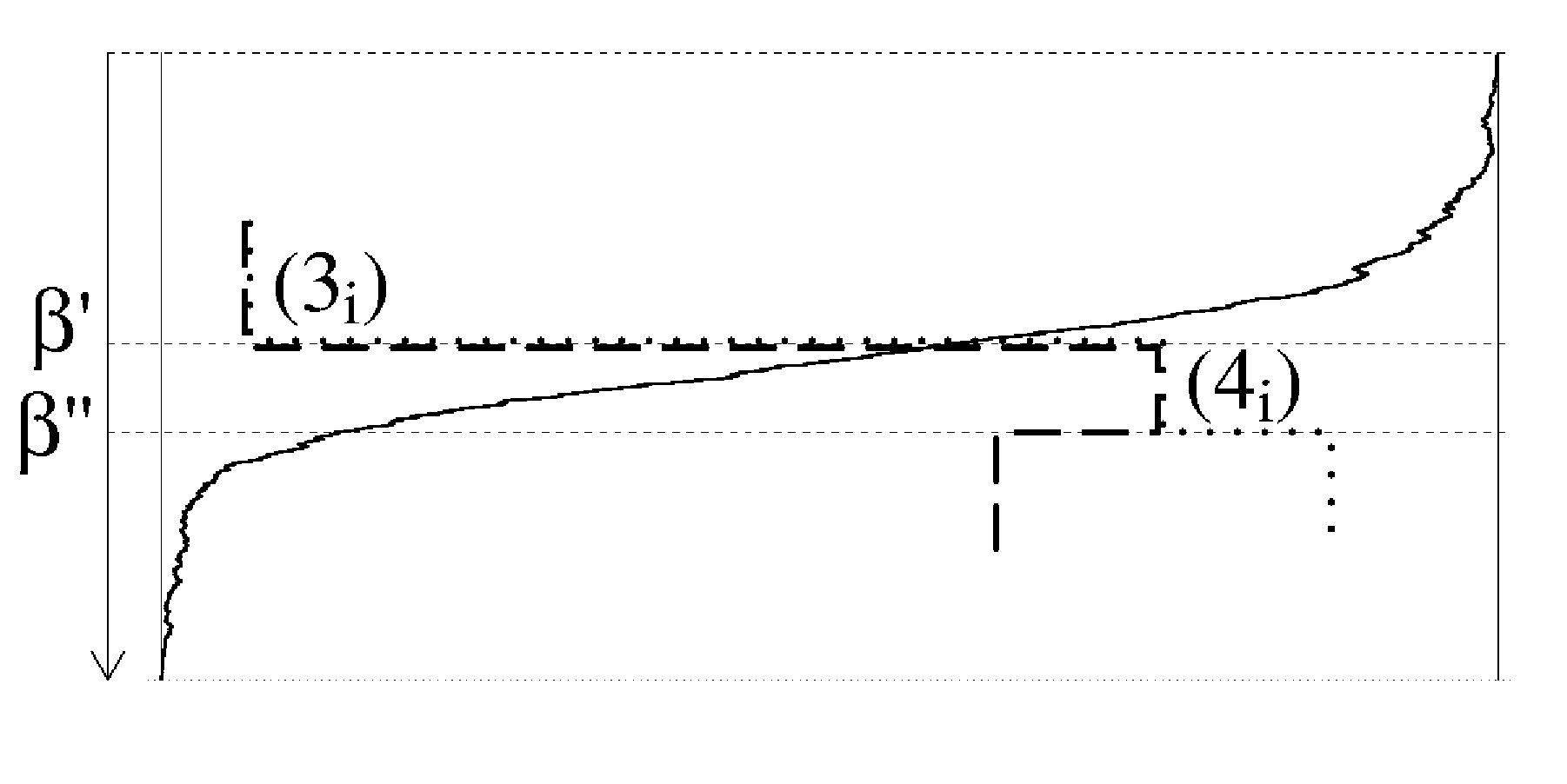}\hspace{1cm}
    \includegraphics[width=7cm]{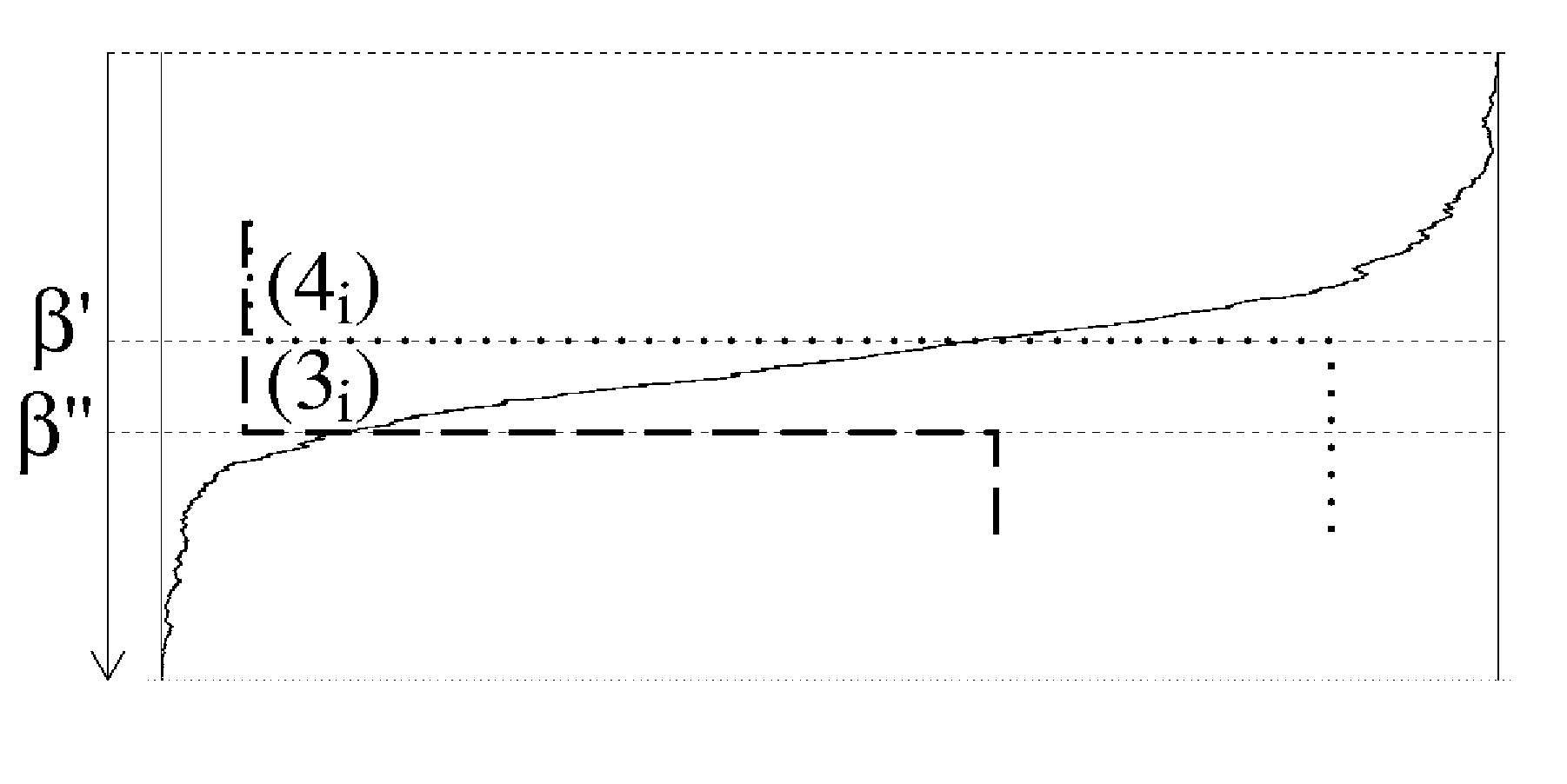}
  \end{center}
  \caption{\label{smallFig}(a) A partition element (a line) is hit by
    an event taking both the $L$- and the $R$-locus to the wild-type background
    at time $\beta'$. Afterwards, at time $\beta''$ the line is split
    in the wild-type background. (b) Here, the $R$-locus is taken to
    the wild-type background at time $\beta'$. Afterwords the
    $L$-locus is taken to the same background at time $\beta''$. The
    outcome is the same. The line moves from the beneficial to the
    wild-type background and is split there.}
  \end{figure}

  Observe that the two-step transitions for the pair
  $\big($\eqref{eq:trans3}, \eqref{eq:trans4b}$\big)$ (see Figure
  \ref{smallFig}(a)) as well as for the pair
  $\big($\eqref{eq:trans4a}, \eqref{eq:trans3}$\big)$ (see Figure
  \ref{smallFig}(b)) are given by
  \begin{equation*} 
  \left(\zeta^B,\zeta^b\right) \longrightarrow \left(\zeta^B \setminus \zeta_{(j)}, \;
    \zeta^b  \cup \{\zeta_{(j)} \cap
    \dickm{\ell},\zeta_{(j)}\cap\dickm{r}\}\right),
  \end{equation*}
  i.e, the partition element both moves from $\zeta^B$ to $\zeta^b$
  and is split in its $L$- and $R$-loci. This proves
  \eqref{eq:claim2}.
\end{step}

\begin{step}(Probable order of events)\\
  Define $\varepsilon:=\frac{(\log \alpha)^{2}}{\alpha}$ and
  $T_{\varepsilon}:= \min\{t\geq 0: X_{t}= \varepsilon \}$. We will
  show that (i) no coalescences, i.e., events $(\mathfrak 1)$, occur
  in $[T_\varepsilon, T]$, (ii) no splits in the beneficial
  background, i.e., events $(\mathfrak 5)$, occur during
  $[0,T_\varepsilon]$ and (iii) splits in the beneficial background,
  i.e., events $(\mathfrak 5)$ do not overlap with other recombination
  events $(\mathfrak 3), (\mathfrak 4)$ with high probability. More
  precisely, we claim
  \begin{align}
    \mathbb P[\mathcal{T}_{\mathfrak 1} \cap \left[ T_{\varepsilon}, T\right]
    \neq \varnothing ] = \mathcal O\Big(
    \frac{1}{(\log\alpha)^2}\Big),
    \label{eq:step3a}
    \\
    \mathbb{P}\left[ \mathcal{T}_{\mathfrak 5} \cap \left[0,
        T_{\varepsilon}\right] \neq \varnothing \right] =
    \mathcal{O}\left(\frac{(\log\alpha)^{2}}{ \alpha}\right), \label{eq:step3b}\\
    \mathbb{P} \left[ \min\mathcal{T}_{\mathfrak 5} <
      \max(\mathcal{T}_{\mathfrak 3}\cup\mathcal{T}_{\mathfrak 4})
    \right] =\mathcal{O}\left(\frac{1}{(\log \alpha)^{2}}.
    \right)\label{eq:step3c}
  \end{align}

  First, \eqref{eq:step3a} coincides with the assertion of Lemma 4.3
  in \cite{EtheridgePfaffelhuberWakolbinger2006}. Second, for
  \eqref{eq:step3b}, we have $X_{t}\leq \frac{(\log
    \alpha)^{2}}{\alpha}$ for all $t \leq T_{\varepsilon}$. Hence we
  get
  \begin{eqnarray*}
    \mathbb{P}\left[\mathcal{T}_{\mathfrak 5} \cap \left[0, T_{\varepsilon}\right] = 
      \varnothing \right]&=& \mathbb{E}\left[\exp\left(- r
        \rho_{LR}\int_{0}^{T_{\varepsilon}}X_{s}ds \right) \right] \\ 
    &\geq& \mathbb{E}\left[ \exp\left(- r \rho_{LR}\;\varepsilon\; T_{\varepsilon} \right)\right]
    \geq  \exp\left(- r \rho_{LR}\;\varepsilon\; \mathbb{E}\left[ T\right] \right).
  \end{eqnarray*}
  By \eqref{eq:T} we see that $\mathbb{E}\left[ T\right]=\frac{2\log
    \alpha}{\alpha}+\mathcal{O}\left(\frac{1}{\alpha}\right)$. By the
  choice of $\varepsilon$, this finally gives
     \begin{equation} \nonumber \mathbb{P}\left[\mathcal{T}_{\mathfrak 5} \cap
      \left[0, T_{\varepsilon} \right] = \varnothing \right] \geq
    1-\mathcal{O}\left(\frac{(\log\alpha)^{2}}{ \alpha}\right).
\end{equation} 
Third, for \eqref{eq:step3c} we write, using $\rho = \mathcal O\left(
  \frac{\alpha}{\log\alpha}\right)$, which might change from
occurrence to occurrence,
\begin{equation}\label{eq:green1}
\begin{aligned}
  \mathbb{P} \left[ \min\mathcal{T}_{\mathfrak 5}  \right. &<
    \max(\mathcal{T}_{\mathfrak 3}\cup\mathcal{T}_{\mathfrak 4}) \left. \right] 
   = 
\\ 
  & = \mathbb{E}\left[ \int_{0}^{T} \mathbb{P}\left[
      \mathcal{T}_{\mathfrak 5} \cap \left[0, t \right] \neq \varnothing
      \big|\max(\mathcal{T}_{\mathfrak 3} \cup \mathcal{T}_{\mathfrak 4}) \in dt, \mathcal
     X \right] \cdot \mathbb{P}\left[\max(\mathcal{T}_{\mathfrak 3}\cup
      \mathcal{T}_{4}) \in dt \big| \mathcal X \right] \right] \\&
  \leq \mathbb{E} \left[ \int_{0}^{T} \left(1- \exp
      \left(-\int_{0}^{t} \rho X_{s}ds\right) \right)\cdot \rho
    (1-X_{t}) \exp\left(-\int_{t}^{T} \rho
      (1-X_{s})ds\right) \right]\\
  & \leq \rho^2 \cdot \mathbb{E}\left[\int_{0}^{T} (1-X_{t})
    \int_{0}^{t} X_{s} dsdt\right]. 
\end{aligned}
\end{equation}
The last term can be estimated using the Green function for the
diffusion \eqref{eq:SDE}. As the right hand side of \eqref{eq:green1}
coincides with the second line of (4.5) in
\cite{EtheridgePfaffelhuberWakolbinger2006} we immediately obtain
\eqref{eq:step3c}.
\end{step}

\medskip

In the next three steps we will show that realizing the different
splits independently from a fixed sample path $\mathcal{X}=
(X_{t})_{0Ê\leq t \leq T}$ will cause only a small error. To see this
we will establish a general result on mixed Poisson processes in Step
4 and apply it to the Poisson processes introduced in Step 2. The
proof of Proposition \ref{PropFirst} will then be concluded by an
application of these two steps.

\begin{step}(General approximations of mixed Poisson processes) \\
  Let $\{\Psi(\delta): \delta > 0\}$, $\{\Phi(\delta): \delta > 0 \}$
  be families of random variables taking values in $\mathbb{R}^{+}$.
  Assume that the expectations $\mathbb{E}[\Psi(\delta)]$,
  $\mathbb{E}[\Phi(\delta)]$ are bounded in $\delta$ and
\begin{equation} \label{var}
 \mathbb{V}[\Psi(\delta)],\mathbb{V}[\Phi(\delta)] = \mathcal{O}\left(\delta \right)
\end{equation}
as $\delta \rightarrow 0$. Denote the distribution function of the
Poisson distribution with parameter $\lambda$ by
$\text{Poi}_{\lambda}(\cdot)$. We claim that for $k, l \in
\mathbb{N}_0$
\begin{eqnarray}\label{approx1}
  \mathbb{E}\left[\text{Poi}_{\Psi(\delta)}(k) \right]&= &
  \text{Poi}_{\mathbb{E}[\Psi(\delta)]}(k)+ \mathcal{O}\left(\delta \right)
  \\ \label{approx2}
  \mathbb{E}\left[\text{Poi}_{\Psi(\delta)}(k) \cdot 
    \text{Poi}_{\Phi(\delta)}(l) \right]&=&\mathbb{E}
  \left[\text{Poi}_{\Psi(\delta)}(k)\right]\cdot 
  \mathbb{E}\left[\text{Poi}_{\Phi(\delta)}(l)\right] + 
  \mathcal{O}\left(\delta  \right)\end{eqnarray}

Note that by a Taylor series approximation, for a random variable
$\Psi$ in $\mathbb R_+$ with second moments and some $\tilde\Psi$
satisfying $\left|\tilde \Psi - \mathbb{E}[\Psi]\right|\leq \left|\Psi -
\mathbb{E}[\Psi]\right|$,
\begin{eqnarray} \nonumber\left| \mathbb{E} \left[ e^{- \Psi}
  \frac{\Psi^{k}}{k!} \right] - e^{- \mathbb{E}[\Psi
    ]}\frac{\mathbb{E}[\Psi ]^{k}}{k!} \right| &=& \left|
  \left[\frac{d^2}{d\Psi^2} \left(e^{- \Psi}
  \frac{\Psi^{k}}{k!}\right)\right]_{\Psi=\mathbb{E}[\Psi]}\right|\cdot
  \mathbb{E} \left[(\tilde \Psi - \mathbb{E}[\Psi])^{2} \right] \\
  \nonumber &\leq&
  e^{- \mathbb{E}\left[\Psi \right]} \left|\left\{ \frac{\mathbb{E}\left[\Psi \right]^{k-2}}{(k-2)!}- 2\frac{\mathbb{E}\left[\Psi \right]^{k-1}}{(k-1)!} + \frac{\mathbb{E}\left[\Psi \right]^{k}}{k!}\right\}\right|\cdot \mathbb{V}\left[\Psi\right] \\
    &\leq& 2 \mathbb{V}\left[\Psi
    \right] \label{Taylor}
\end{eqnarray} 
where the terms in $\{ \ldots \}$ only show up if the denominators are
non-zero and the last step follows from the fact that the Poisson
weights in $\{ \ldots \}$ lie in $[0,1]$. As this holds for every
$\Psi(\delta)$, (\ref{approx1}) follows immediately from (\ref{var}).
Moreover, by a calculation similar to (\ref{Taylor}),
\begin{eqnarray*}
  \mathbb{V} \left[\text{Poi}_{\Psi(\delta)}(k) \right] 
  = \mathbb{E}\left[ e^{-2\Psi(\delta)} \frac{\Psi(\delta)^{2k}}{(k!)^{2}}\right] - \mathbb{E}\left[e^{-\Psi(\delta)}Ê\frac{\Psi(\delta)^{k}}{k!} \right]^{2}
  =\mathcal O \big(\mathbb{V}\left[ \Psi(\delta) \right]\big) 
  = \mathcal{O}\left(\delta \right).
\end{eqnarray*}
Additionally, (\ref{approx2}) follows easily from the fact that
\begin{eqnarray*}
  &&\big| \mathbb{E}\left[\text{Poi}_{\Psi(\delta)}(k) \cdot \text{Poi}_{\Phi(\delta)}(l) \right]-\mathbb{E}\left[\text{Poi}_{\Psi(\delta)}(k)\right]\cdot \mathbb{E}\left[\text{Poi}_{\Phi(\delta)}(l)\right] \big| \\
  &&\qquad \qquad  =  \big| \text{Cov}\left[\text{Poi}_{\Psi(\delta)}(k) \cdot \text{Poi}_{\Phi(\delta)}(l) \right] \big| 
  \leq \sqrt{
    \mathbb{V}\left[ \text{Poi}_{\Psi(\delta)}(k)\right] \cdot \mathbb{V}\left[\text{Poi}_{\Phi(\delta)}(l) \right]} = \mathcal{O}\left( \delta \right)
\end{eqnarray*}
by the Cauchy-Schwarz inequality.
\end{step}

\begin{step}(Green function estimates)\\
  Set $\rho=\gamma \frac{\alpha}{\log \alpha}$ where $\gamma =
  \gamma_{LR}$ for geometry (i) and $\gamma= \gamma_{LS}+ \gamma_{SR}$
  for geometry (ii). Using our approximations from Step 4 we will show
  next
\begin{eqnarray} \label{Poi1}
\mathbb{P}\left[|\mathcal{T}_{\mathfrak 5}|= k \right] &=&\text{Poi}_{\mathbb{E}[r\rho\int_{0}^{T}X_{s}ds]}(k)+ \mathcal{O}\left( \frac{1}{(\log \alpha)^{2}}\right) \\  \nonumber
\mathbb{P} \left[\big|(\mathcal{T}_{\mathfrak 3} \cup \mathcal{T}_{\mathfrak 4}) \cap [T_{\varepsilon}, T]\big|= k,  |\mathcal{T}_{\mathfrak 5}|= l \right]
&=&\mathbb{P}\left[\big|(\mathcal{T}_{\mathfrak 3} \cup \mathcal{T}_{\mathfrak 4})\cap [T_{\varepsilon}, T]\big|= k \right] \cdot \mathbb{P}\left[|\mathcal{T}_{\mathfrak 5}|= l\ \right] \\ \label{Poi2}
&&  \qquad \qquad \qquad \qquad + \;\mathcal{O}\left(\frac{1}{(\log \alpha)^{2}} \right)
\end{eqnarray}
as $\alpha \rightarrow \infty$.
To see this, set $\delta= \frac{1}{(\log \alpha)^{2}}$ and define
\[
 \Psi(\delta)= r\rho\int_{0}^{T} X_{s}ds, \qquad\qquad \Phi(\delta)= (\ell+r)\rho \int_{T_{\varepsilon}}^{T}(1-X_{s})ds
\]
Observe that for $k= 0,1,2, \ldots$
\begin{align}
\mathbb{P}\left[|\mathcal{T}_{\mathfrak 5}|= k\right]&= \mathbb{E}\left[ \text{Poi}_{\Psi(\delta)}(k)\right]\label{Poi3}\\
\mathbb{P}\left[\big|(\mathcal{T}_{\mathfrak 3} \cup \mathcal{T}_{\mathfrak 4})\cap [T_{\varepsilon}, T]\big|= k \right]& = \mathbb{E}\left[\text{Poi}_{\Phi(\delta)}(k) \right]\nonumber
\end{align}
because $\mathcal{T}_{\mathfrak 3}$, $\mathcal{T}_{\mathfrak 4}$, $\mathcal{T}_{\mathfrak 5}$ are randomly time-changed Poisson processes. By (\ref{approx1}) and (\ref{approx2}), (\ref{Poi1}) and (\ref{Poi2}) follow once we have shown
\begin{eqnarray}
\mathbb{E}\left[\rho \int_{T_{\varepsilon}}^{T}(1-X_{s})ds \right] &\leq& \mathbb{E}\left[\rho \int_{0}^{T}X_{s}ds \right] \leq 2\gamma + \mathcal{O}\left(\frac{1}{\alpha} \right) \label{key1} \\
\mathbb{V}\left[\rho \int_{T_{\varepsilon}}^{T}(1-X_{s})ds \right] &\leq& \mathbb{V}\left[\rho \int_{0}^{T}X_{s}ds \right] = \mathcal{O}\left(\frac{1}{(\log\alpha)^{2}} \right) \label{key2}
\end{eqnarray}
as $\alpha \rightarrow \infty$.\\
First observe that $\left(X_{t} \right)_{0 \leq t \leq T}$ has the
same distribution as $(1-X_{T-t})_{0 \leq t\leq T}$ by
time-reversibility (see e.g. \cite{KarlinTaylor1981,
  Griffiths2003}). Hence the inequalities on the left hand side of
(\ref{key1}) and (\ref{key2}) follow. Second, we verify the
expressions on the right hand side of (\ref{key1}) and (\ref{key2}) by
an application of the Green function $G(.,.)$ of the diffusion
$(X_{t})_{0 \leq t\leq T}$.  This function satisfies
\[
\mathbb E_x\left[ \int_0^T g(X_t) dt\right] = \int_0^1 G(x,y) g(y) dy
\]
where $\mathbb E_x[.]$ refers to the path $(X_t)_{0\leq t\leq T}$ with
$X_0=x$ and $\mathbb E[.] := \mathbb E_0[.]$. The Green function is
given by
\[
G(x, y)=
\begin{cases}
  \frac{\left(1-e^{-\alpha(1-y)}\right) \left(1-e^{-\alpha y} \right)}{\alpha y \left(1-y \right)\left(1-e^{-\alpha} \right)} \qquad &\text{ if } x \leq y \\
  \frac{\left(e^{-\alpha x} -e^{-\alpha} \right)\left(e^{\alpha y}-1
    \right) \left(1-e^{-\alpha y} \right)} {\alpha y \left(1-y \right)
    \left(1-e^{-\alpha} \right) \left(1-e^{-\alpha x} \right)} \qquad
  &\text{ if } x\geq y,
\end{cases}
\]
see e.g. \cite{KarlinTaylor1981,
  EtheridgePfaffelhuberWakolbinger2006}. More generally, $G(.,.)$
satisfies
\begin{eqnarray*}
&&\mathbb{E}_{x}\left[\int_{0}^{T} \int_{t_{1}}^{T} \ldots \int_{t_{k-1}}^{T} g_{k}(X_{t_{k}})\ldots g_{1}(X_{t_{1}})dt_{k}\ldots dt_{1} \right] \\
  &&\qquad  \qquad \qquad \qquad
  = \int_{0}^{1} \ldots \int_{0}^{1} G(x, x_{1}) \ldots G(x_{k-1}, x_{k}) g_{1}(x_{1}) \ldots g_{k}(x_{k})dx_{k}\ldots dx_{1}
\end{eqnarray*}
for all $k=1, 2, \ldots$ which can be proved by induction. We may thus
write, because $G(x,y) = G(0,y)$ for $y\geq x$,
\begin{align*} 
  \mathbb{V}\Big[  \rho\int_{0}^{T}X_{s}ds\Big] &= 
   \rho^2 \left( 2 \int_0^1 \int_0^1 G(0,x) G(x,y) xy dy dx - 2\int_0^1 \int_x^1 G(0,x) G(0,y) xy dy dx\right) \\
  & = 2\rho^2 \int_0^1 \int_0^x G(0,x) G(x,y) xy dy dx
  \: \leq \: 2 \rho^2 \int_0^1 \int_0^x G(0,x) G(x,y) dy dx \\
  & = 2\rho^2 \mathbb V[T] = \mathcal{O}\left(
    \frac{1}{(\log\alpha)^{2}}\right)
\end{align*}
by \eqref{eq:T} which gives (\ref{key2}).
\end{step}

\begin{step}(Approximate independence)\\
  As we have seen in (\ref{eq:claim}) the distribution of
  $\zeta^{\mathcal X}_T$ is determined by the distribution of the
  order of events in the Poisson processes $\mathcal T_{\mathfrak 1}$,
  $\mathcal T_{\mathfrak 3}$, $\mathcal T_{\mathfrak 4}$ and $\mathcal
  T_{\mathfrak 5}$.  The calculations in Step 3 allow us to make the
  assumptions
  \[
  \mathcal{T}_{\mathfrak 1}\cap \left[T_{\varepsilon}, T \right] =
  \varnothing, \qquad \mathcal{T}_{\mathfrak 5}\cap \left[0,
    T_{\varepsilon} \right] = \varnothing,\qquad \max
  (\mathcal{T}_{\mathfrak 3} \cup \mathcal{T}_{\mathfrak 4}) < \min
  \mathcal{T}_{5}
  \]
  on the ordering of events in these Poisson processes as these events
  have probability $1-\mathcal O\big( (\log\alpha)^{-2}\big)$.
  Furthermore, we know from \eqref{eq:claim2} that events in $\mathcal
  T_{\mathfrak 3}$ and $\mathcal T_{\mathfrak 4}$ may be exchanged
  without changing the distribution of $\zeta^{\mathcal X}_T$. Hence,
  the distribution of $\zeta^{\mathcal X}_T$ is determined once the
  joint distribution of
  $$\mathcal{T}_{\mathfrak 1} \cap \left[ 0, T_{\varepsilon}\right], \qquad 
  \mathcal{T}_{\mathfrak 3}\cap \left[ 0, T_{\varepsilon}\right],
  \qquad \mathcal{T}_{\mathfrak 4}\cap [0,T_\varepsilon], \qquad
  \left|(\mathcal{T}_{\mathfrak 3}\cup \mathcal{T}_{\mathfrak 4}) \cap
    \left[ T_{\varepsilon}, T\right] \right|,\qquad
  \left|\mathcal{T}_{\mathfrak 5}\right| $$ is known. To approximate
  the joint distribution of these objects, define
\[
\mathcal{T}_{\mathfrak i}^{\varepsilon}:= \mathcal{T}_{\mathfrak i}
\cap \left[0, T_{\varepsilon} \right]\text{, } \mathfrak i=
\mathfrak{1,3,4} \quad \text{and} \quad K_{\mathfrak {3,4}}:=
\big|\left(\mathcal{T}_{\mathfrak 3} \cup \mathcal{T}_{\mathfrak 4}
\right) \cap \left[T_{\varepsilon}, T\right] \big| \text{,} \quad K_{
  \mathfrak 5}:=\big| \mathcal{T}_{\mathfrak 5} \big|.
\]
We will prove 
\begin{equation} \label{independence} \mathbb P \circ
  \left(\mathcal{T}_{\mathfrak 1}^{\varepsilon},
    \mathcal{T}_{\mathfrak 3}^{\varepsilon}, \mathcal{T}_{\mathfrak
      4}^{\varepsilon}, K_{\mathfrak {3,4}}, K_{\mathfrak 5}
  \right)^{-1} = \mathbb P\circ\left(\mathcal{T}_{\mathfrak
      1}^{\varepsilon}, \mathcal{T}_{\mathfrak 3}^{\varepsilon},
    \mathcal{T}_{\mathfrak 4}^{\varepsilon}, K_{\mathfrak{3,4}}
  \right)^{-1} \otimes \;
  \text{Poi}_{\mathbb{E}\left[r\rho\int_{0}^{T}X_{s}ds \right]} +
  \mathcal{O}\left(\frac{1}{\left(\log \alpha \right)^{2}} \right)
\end{equation}
where $\mathbb P\circ X^{-1}$ is the image measure of the random
variable $X$ under $\mathbb P$ and the Landau symbol in this context
gives the order in variation distance of the distributions.

Once \eqref{independence} is shown we conclude that $K_{\mathfrak 5}$
is approximately independent of all other events. Furthermore, its
distribution may be interpreted as the sum of $r$ Poisson
distributions with parameter $\mathbb{E}\left[\rho\int_{0}^{T}X_{s}ds
\right]$. These determine the number of split events on all partition
elements $\xi\in\pi$ with $\xi\cap\dickm r\neq \varnothing$. A
partition element splits, if it is hit by at least one split event.
The probability for a split of a partition element is thus given,
using \eqref{Poi1} and \eqref{Poi3} for $k=0$, by
\[
1-\exp\Big( - \rho\cdot \mathbb{E}\left[\int_{0}^{T}X_{s}ds
\right]\Big) = 1 - \mathbb E\Big[ \exp\Big( - \rho\int_0^T X_s
ds\Big)\Big] + \mathcal O\Big( \frac{1}{(\log\alpha)^2}\Big).
\]
with $\rho=\rho_{LR}$ for geometry (i) and $\rho=\rho_{LS} +
\rho_{SR}$ for geometry (ii). Observe that $\Gamma_\pi$ is determined
by the distribution of $\left(\mathcal{T}_{\mathfrak 1}^{\varepsilon},
  \mathcal{T}_{\mathfrak 3}^{\varepsilon}, \mathcal{T}_{\mathfrak
    4}^{\varepsilon}, K_{\mathfrak {3,4}}\right)$ if $K_{\mathfrak 5}$
is known. The random partition $\Delta_\pi$ is determined by the
distribution of $\left(\mathcal{T}_{\mathfrak 1}^{\varepsilon},
  \mathcal{T}_{\mathfrak 3}^{\varepsilon}, \mathcal{T}_{\mathfrak
    4}^{\varepsilon}, K_{\mathfrak {3,4}}\right)$ independently of
$K_{\mathfrak 5}$. So, Proposition \ref{PropFirst} is a consequence of
the approximate independence of $\left(\mathcal{T}_{\mathfrak
    1}^{\varepsilon}, \mathcal{T}_{\mathfrak 3}^{\varepsilon},
  \mathcal{T}_{\mathfrak 4}^{\varepsilon}, K_{\mathfrak {3,4}}\right)$
and $K_{\mathfrak 5}$ given by \eqref{independence}.

\smallskip

We write
\begin{align*}
  \mathbb P \circ \big(\mathcal{T}_{\mathfrak 1}^{\varepsilon}, &
  \mathcal{T}_{\mathfrak 3}^{\varepsilon}, \mathcal{T}_{\mathfrak
    4}^{\varepsilon}, K_{\mathfrak{3,4}}, K_{\mathfrak{5}} \big)^{-1}
  = \int \mathbb{P}_{\mathcal X} \circ \left(\mathcal{T}_{\mathfrak
      1}^{\varepsilon}, \mathcal{T}_{\mathfrak 3}^{\varepsilon},
    \mathcal{T}_{\mathfrak 4}^{\varepsilon}, K_{\mathfrak {3,4}},
    K_{\mathfrak{5}} \right)^{-1}\;\mathbb{P}\left[ d\mathcal{X}
  \right] \\ & = \int \mathbb{P}_{(X_{t})_{0\leq t \leq T^{\varepsilon}}}
  \circ \left(\mathcal{T}_{\mathfrak 1}^{\varepsilon},
    \mathcal{T}_{\mathfrak 3}^{\varepsilon}, \mathcal{T}_{\mathfrak
      4}^{\varepsilon} \right)^{-1} \mathbb{P}\left[d(X_{t})_{0\leq t
      \leq T^{\varepsilon}}\right]\\ & \qquad\qquad \otimes \int
  \mathbb{P}_{(X_{t})_{T^{\varepsilon}\leq t \leq T}}\circ \left(K_{\mathfrak
      {3,4}}, K_{\mathfrak 5} \right)^{-1}
  \mathbb{P}\left[d(X_{t})_{T^{\varepsilon}\leq t \leq T}\right] +
  \mathcal{O}\left(\frac{(\log \alpha)^{2}}{\alpha} \right)
\end{align*}
where we have used the fact that $T_{\varepsilon}$ is a stopping time and
the strong Markov property of the process $\mathcal X$. Note that by
\eqref{eq:step3b} we may assume $K_{ \mathfrak
  5}=\big|\mathcal{T}_{\mathfrak 5} \cap \left[T_{\varepsilon}, T \right]
\big|$ which gives an error of $\mathcal{O}\left(\frac{(\log
    \alpha)^{2}}{\alpha} \right)$ in probability. From Steps 4 and 5
we get
\begin{multline*}
  \int \mathbb{P}_{(X_{t})_{T^{\varepsilon}\leq t \leq T}}\circ
  \left(K_{\mathfrak{3,4}}, K_{\mathfrak{5}} \right)^{-1}
  \mathbb{P}\left[d(X_{t})_{T^{\varepsilon}\leq t \leq T}\right]
  \\
  =
  \text{Poi}_{\mathbb{E}\left[(\ell+r)\rho\int_{T_{\varepsilon}}^{T}\left(1-X_{s}
      \right)ds \right]}\; \otimes\;
  \text{Poi}_{\mathbb{E}\left[r\rho\int_{T_{\varepsilon}}^{T}X_{s}ds
    \right]} + \mathcal{O}\left(\frac{1}{(\log\alpha)^{2}}\right)
\end{multline*}
Rewriting
$$\text{Poi}_{\mathbb{E}\left[(\ell+r)\rho\int_{T_{\varepsilon}}^{T}\left(1-X_{s}
    \right)ds \right]} = \int \mathbb{P}_{(X_{t})_{T^{\varepsilon}}\leq t
  \leq T}\circ\big(K_{
  \mathfrak{3,4}}\big)^{-1}\mathbb{P}\left[d(X_{t})_{T^{\varepsilon}\leq
    t \leq T}\right] ,$$ and using the strong Markov property of
$\mathcal X$ a second time we get
\begin{align*}
  \mathbb P\circ \left(\mathcal{T}_{\mathfrak 1}^{\varepsilon},
    \mathcal{T}_{\mathfrak 3}^{\varepsilon}, \mathcal{T}_{\mathfrak
      4}^{\varepsilon}, K_{\mathfrak{3,4}}, K_{\mathfrak{5}}
  \right)^{-1} &= \int \mathbb{P}_{(X_{t})_{0\leq t \leq
      T^{\varepsilon}}}\circ \left(\mathcal{T}_{\mathfrak
      1}^{\varepsilon}, \mathcal{T}_{\mathfrak 3}^{\varepsilon},
    \mathcal{T}_{\mathfrak 4}^{\varepsilon}
  \right)^{-1}\mathbb{P}\left[d(X_{t})_{0\leq t \leq
      T^{\varepsilon}}\right]
  \\
  & \qquad \otimes \int \mathbb{P}_{(X_{t})_{T^{\varepsilon}\leq t \leq
      T}} \circ \left(K_{\mathfrak{3,4}}\right)^{-1}
  \mathbb{P}\left[d(X_{t})_{T^{\varepsilon}\leq t
      \leq T}\right] \\
  & \qquad \qquad \qquad \otimes \quad
  \text{Poi}_{\mathbb{E}\left[r\rho\int_{0}^{T}X_{s}ds \right]} +
  \mathcal{O}\left(\frac{1}{(\log\alpha)^{2}}\right)\\ &= \mathbb P
  \circ \left(\mathcal{T}_{\mathfrak 1}^{\varepsilon},
    \mathcal{T}_{\mathfrak 3}^{\varepsilon}, \mathcal{T}_{\mathfrak
      4}^{\varepsilon}, K_{\mathfrak {3,4}} \right)^{-1} \otimes
  \text{Poi}_{\mathbb{E}\left[r\rho\int_{0}^{T}X_{s}ds \right]} +
  \mathcal{O}\left(\frac{1}{\left(\log \alpha \right)^{2}} \right)
\end{align*}
and we are done.
\end{step}
\end{proof}

By Proposition \ref{PropFirst}, events $(5)$ can be generated
independently of the frequency path and of all other events.  The
rates of the recombination events $(3), (4), (6_i)$ at time $\beta$
are all proportional to $(1-X_{T-\beta})$.  This is reminiscent of the
case of only one neutral locus, studied in
\cite{EtheridgePfaffelhuberWakolbinger2006}, where a line carrying one
neutral locus in recombination distance $\rho$ recombines to the
wild-type background with rate $\rho(1-X_{T-\beta})$. As a consequence we can use
the same techniques used there, especially their Proposition 3.6.
which states that a marked Yule tree approximately gives the same
partition as the structured coalescent.

\begin{definition}\label{def:4}
  Define a $\mathcal P'_{\dickm \ell\,\cup\,\dickm r}$-valued random
  variable $\Xi_\pi$ as follows: For all partition elements
  $\xi\in\pi$ which $\xi\cap\dickm\ell \neq \varnothing, \xi\cap\dickm
  r\neq\varnothing$, i.e., $\xi$ carries both left and right loci,
  split the partition element in its left and right loci,
  $\xi\cap\dickm\ell, \xi\cap\dickm r$ according to \eqref{Split}.
  Denote
  the resulting partition by $\pi'$.
  
  Let $\mathbf Y$ be an infinite Yule tree with branching rate
  $\alpha$. Moreover, consider the random tree $\mathbf Y_{|\pi'|}$
  which arises by sampling $|\pi'|$ lines from $\mathbf Y$ at
  infinity. Identify each of the $|\pi'|$ partition elements of $\pi'$
  with one sampled line. Between the root of the Yule tree $\mathbf Y$
  starts and the time it has $\lfloor 2\alpha \rfloor$ lines, mark all
  lines by the following procedure:

  For geometry (i), the tree is marked by Poisson processes with rates
  $\rho_{SL}$ and $\rho_{LR}$. These marks are relabelled such that
  each branch is hit by at most one mark. Call the corresponding marks
  $SL$-, $LR$- and $SLR$-marks. The following rules are applied:
  \begin{enumerate}
  \item[(a)] If the Poisson process with rate $\rho_{SL}$ puts the first
    (backward in time) mark at time $t$ from the root, start a Poisson
    process with rate $\rho_{LR}$ and run it for time $t$. If an event
    occurs during this time, the branch is marked by an $SLR$-mark,
    otherwise by an $SL$-mark.
  \item[(b)] If the Poisson process with rate $\rho_{LR}$ puts the
    first (backward in time) mark distinguish the following two cases:
    if the Poisson process with rate $\rho_{SL}$ hits the branch as
    well, it obtains an $SLR$-mark. Otherwise, it obtains an
    $LR$-mark.
  \end{enumerate}

  For geometry (ii), mark the tree by two independent Poisson
  processes with rates $\rho_{LS}$ and $\rho_{SR}$. If a branch is hit
  by one or more events of the Poisson process with rate $\rho_{LS}$,
  it gets an $LS$-mark. If it is hit by one or more events with rate
  $\rho_{SR}$, it additionally gets an $SR$-mark. 

  The result of this procedure is a marked Yule tree $\mathbf
  Y_{|\pi'|}$. Given $\pi'$ and the marked Yule tree $\mathbf
  Y_{|\pi'|}$ we use the same equivalence relation as given in
  \eqref{eq:equivGeoi} and \eqref{eq:equivGeoii} to define
  $\pi''\in\mathbf P'_{\dickm\ell \cup \dickm r}$. Furthermore, we
  use \eqref{eq:uni} and \eqref{eq:unii} to define the random
  partition
  $$ \Xi_\pi:= ( \{\pi_f''\}, \pi''\setminus \{\pi_f''\}).$$ 
\end{definition}

\begin{figure}
\hspace{3cm} (a) \hspace{7cm}(b)

\begin{center}
\includegraphics[width=7cm]{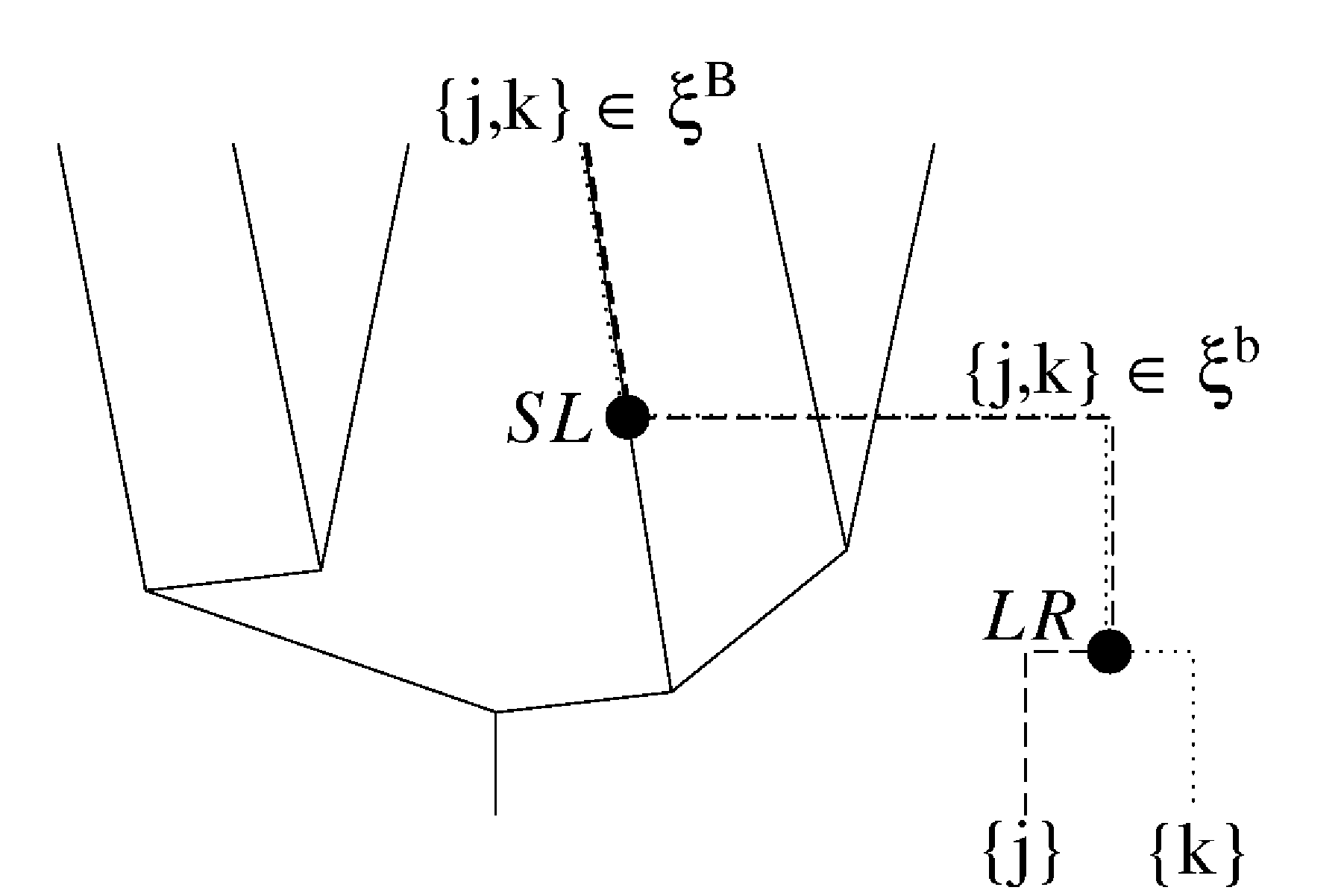} \hspace{0.5cm}
\includegraphics[width=7cm]{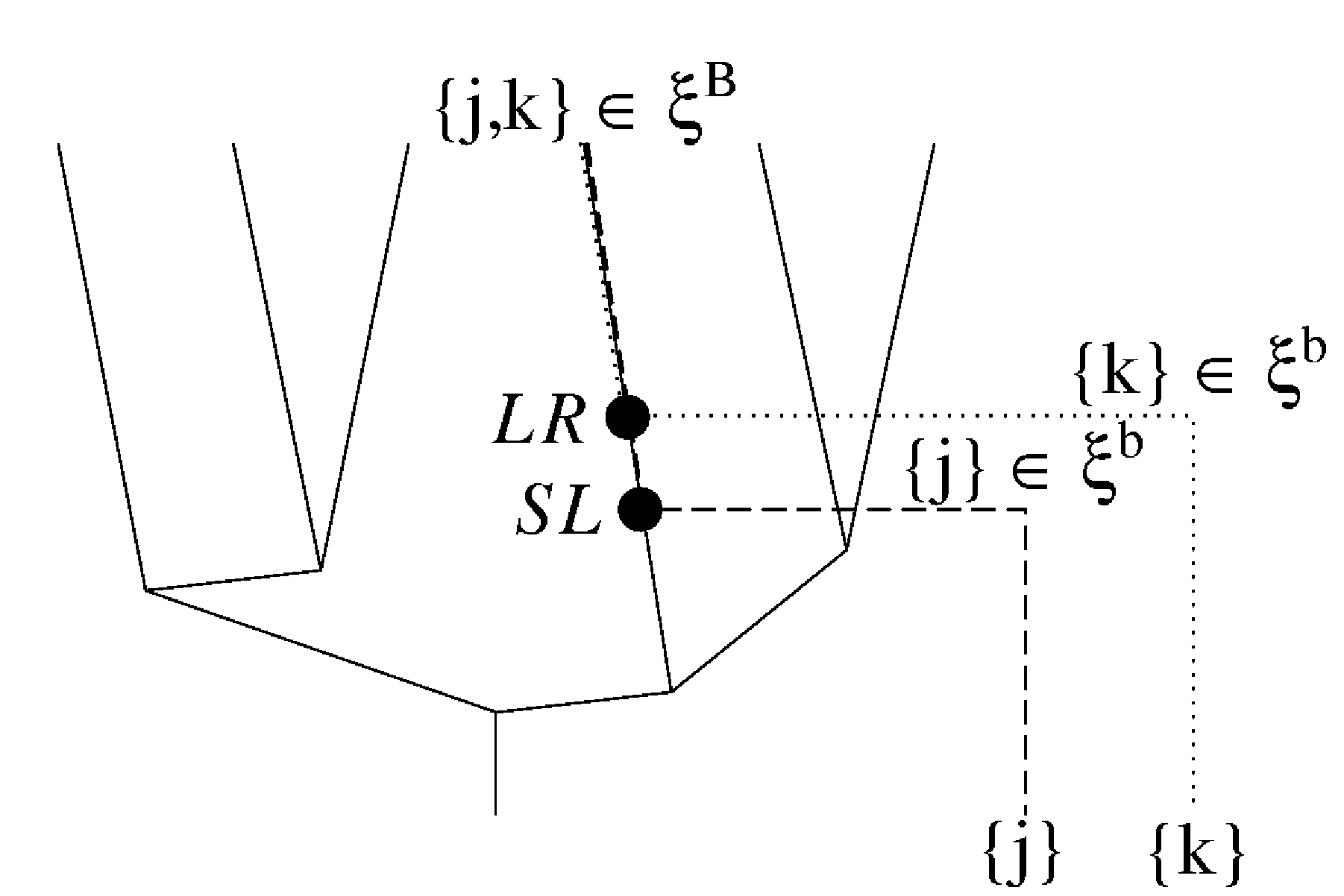}
\end{center}
\caption{\label{smallYule}
There are two possibilities how an $SLR$-mark may occur. Here, $SL$
and $LR$ refer to points in the Poisson processes with rates
$\rho_{SL}$ and $\rho_{LR}$.  See text for further explanation.
}
\end{figure}

\begin{example} The two cases in which an $SLR$-mark occurs for
  geometry (i) are illustrated in Figure \ref{smallYule}. Consider the
  line in the sample Yule tree which can be identified with the
  partition element $\{j,k\}$ where $j\in\dickm \ell$ and $k\in\dickm
  r$. Consider case (a) first, shown on the left side of Figure
  \ref{smallYule}: The $SL$-mark hitting a branch in $\mathbf
  Y_{|\pi'|}$ leads to a jump of the partition element into the
  wild-type background. We now have to consider the additional Poisson
  process at rate $\rho_{LR}$ to determine whether or not the line
  will split within the wild-type background. If an event with rate
  $\rho_{LR}$ occurs, the $L$- is separated from the $R$-locus on this
  line. Case (b) is illustrated on the right side of Figure
  \ref{smallYule}. Here, the line which refers to the partition
  element $\{j,k\}$ is first (backward in time) hit by an $LR$-mark,
  bringing the $R$-locus into the wild-type background, and after that
  an additional $SL$-mark hits the same branch, which additionally
  brings the $L$-locus into the wild-type background.
  In both cases the loci $j$ and $k$ end up separated in the wild-type
  background. This is summarized in Definition \ref{def:4} by an $SLR$-mark.
\end{example}

\noindent
As a next step in the Proof of Theorem \ref{T} we now show that
$\Delta_\pi \approx \Xi_\pi$.

\begin{proposition}\label{PropSecond}
  Let $\pi\in\mathcal P'_{\dickm \ell\cup\dickm r}$ and $\Delta_\pi$
  and $\Xi_\pi$ be as in Definitions \ref{def:3} and \ref{def:4}.
  Then,
  $$ \sup_{\xi\in\mathcal P'_{\dickm \ell\cup\dickm r}} 
  \big|\mathbb P[\Delta_\pi = \xi] - \mathbb P[\Xi_\pi=\xi] \big|
  = \mathcal O\Big( \frac{1}{(\log\alpha)^2}\Big). $$
\end{proposition}

\begin{proof}
  As the mechanism to generate splits in the beneficial background is
  the same for both random partitions, $\Delta_\pi$ and $\Xi_\pi$,
  we concentrate on all other events.

  The proof follows along the lines of the Yule approximation in the
  case of only one neutral locus, given in \cite[Definition 3.3. and
  Section 4.3.]{EtheridgePfaffelhuberWakolbinger2006}. The crucial
  observation is that by a random time change $t\mapsto\tau$ given by
  $d\tau = (1-X_t)dt$ the frequency path $\mathcal X$, given by
  \eqref{eq:SDE}, is taken to the solution $\mathcal Z = (Z_t)_{t\geq
    0}$ of
  \begin{align} \label{eq:timechange} dZ = \alpha Z \coth(\alpha Z)dt
    + \sqrt{Z} dW
  \end{align}
  with a standard Brownian motion $W$ and $Z_0=0$. This is an
  $\alpha$-supercritical Feller branching process conditioned on
  non-extinction.  It was shown in \cite{EvansOConnell1994} and
  \cite{OConnell1993} that the genealogy of the $\alpha$-supercritical
  branching process is a Yule process with branching rate $\alpha$.
  Observe that the time-transformation $t\mapsto\tau$ only works until
  the supercritical branching process has reached frequency 1. From
  4.5(b) in \cite{EtheridgePfaffelhuberWakolbinger2006} we see that at
  this time the number of lines in the Yule process is Poisson
  distributed with mean $2\alpha$. (The additional factor of 2 arises
  because we made the assumption that the individual offspring
  variance in the underlying Cannings model is 1 rather than 2. See
  also \cite{PfaffelhuberHauboldWakolbinger2006}.) However, as typical
  deviations in this Poisson distribution are of the order
  $\sqrt\alpha\ll\alpha$ we may instead assume that the Yule process
  has $\lfloor 2\alpha \rfloor$ lines.  This was made precise in the
  proof of Proposition 4.7. in
  \cite{EtheridgePfaffelhuberWakolbinger2006}.

  Moreover, for geometries (i) and (ii) the rates in the process $\xi$
  change at time $\beta$ from $\rho_{SL}(1-X_{T-\beta})$,
  $\rho_{LR}(1-X_{T-\beta})$ to $\rho_{SL}$, $\rho_{LR}$ and from
  $\rho_{LS}(1-X_{T-\beta})$, $\rho_{SR}(1-X_{T-\beta})$ to
  $\rho_{LS}$, $\rho_{SR}$, respectively. Especially, the time-changed
  rates are constant.  Under the random time change the coalescence
  rate (1) changes at time $\beta$ from $1/X_{T-\beta}$ to
  $1/(X_{T-\beta}(1-X_{T-\beta}))$. However, it was shown in
  \cite[Proposition 4.2.]{EtheridgePfaffelhuberWakolbinger2006} that
  the change of these rates can only produce an error in probability
  of order $\mathcal O\big((\log\alpha)^{-2}\big)$. This fact was used
  in \cite[Lemma 4.5., Proposition
  4.7.]{EtheridgePfaffelhuberWakolbinger2006} to prove that the marked
  Yule process gives an accurate approximation in the case for one
  neutral locus. However, this result carries over to the present
  situation because all Poisson processes along the Yule process have
  constant rates.

  It remains to check whether the equivalence relation $\Xi_\pi$
  coincides with $\Delta_\pi$ given the change in the coalescence rate
  has no effect. First of all, realize the splits in the beneficial
  background according to Definition \ref{def:3}. Then, take $j,k\in
  \dickm\ell \cup \dickm r$ and trace their partition elements
  backwards up to time $t=0, \beta=T$. We only consider geometry (i)
  and $j\in\dickm\ell, k\in\dickm r$, since the other cases
  $j,k\in\dickm\ell$ and $j,k\in\dickm r$ and all cases for geometry
  (ii) are similar. If we consider the process $\eta^{\mathcal X}$
  from Definition \ref{def:3} without any recombination events we
  would obtain a tree $\Y$ for the genealogy relating $j$ and $k$.
  However, recombination events may cause the $L$-locus $j$ and the
  $R$-locus $k$ to end up in different partition element in the random
  partitions $\Delta_\pi$. This will be the case if and only if one of
  the following events occurs in the process $\eta^{\mathcal X}$:
\begin{itemize}
\item[(a)] a recombination event $(3_i)$ with rate
  $\rho_{SL}\left(1-X_{}\right)$ on
  $\!\!\!\!\!\!\!\!\Yup\!\!\!\!\!\!\!\!$, which takes either $j$ or
  $k$ to the wild-type background before coalescence,
\item[(b)] a recombination event $(4_i)$ with rate
  $\rho_{LR}\left(1-X_{}\right)$ on
  $\!\!\!\!\!\!\!\!\Yupri\!\!\!\!\!\!\!\!$, which takes $k$ to the
  wild-type background before coalescence with $j$,
\item[(c)] an event $(4_i)$ with rate $\rho_{LR}\left(1-X_{}\right)$ on
  $\!\!\!\!\!\!\!\!\Ybottom\!\!\!\!\!\!\!\!$ before (backward in time)
  an event with rate $\rho_{SL}\left(1-X_{}\right)$ happens on that
  branch; in this case $j$ and $k$ have coalesced, but a recombination
  event brings $k$ to the wild-type background without $j$,
\item[(d)] an event $(3_i)$ with rate $\rho_{SL}\left(1-X_{}\right)$
  on $\!\!\!\!\!\!\!\!\Ybottom\!\!\!\!\!\!\!\!$ before (backward in
  time) an event with rate $\rho_{LR}\left(1-X_{}\right)$ happens on
  that branch, which brings both $j$ and $k$ to the wild-type
  background.  Here, an event $(6_i)$ at rate $\rho_{LR}(1-X_{})$ happens
  which splits $j$ and $k$ in the wild-type background.
\end{itemize}
The trees in events (a)-(d) refer to trees generated by
$\eta^{\mathcal X}$. By the random time change and our assumption that
the change in coalescence rate does not alter random partitions we can
as well take trees generated by the Yule process and change the rates
$\rho_{SL}(1-X)$ and $\rho_{LR}(1-X)$ to $\rho_{SL}$ and $\rho_{LR}$.
Hence we are dealing with a Yule tree with branching rates $\alpha$
marked by Poisson processes with rates $\rho_{SL}$ and $\rho_{LR}$
which is the exact situation of Definition \ref{def:4}. Using the
definition of the $SL$-, $LR$- and $SLR$-marks, we note that
\begin{itemize}
\item (a) produces either an $SL$- or an $SLR$-mark on $\Yup$,
\item (b) produces an $LR$-mark on $\Yupri$,
\item (c) and (d) produce either an $LR$- or an $SLR$-mark on
  $\Ybottom$.
\end{itemize}
If none of these marks occur, $j$ and $k$ are in the same partition
element of $\Xi_\pi$ by \eqref{eq:equivGeoi}. Hence $\Delta_\pi$ and
$\Xi_\pi$ coincide with high probability.

\end{proof}

We conclude the proof of Theorem \ref{T} by showing that $\Xi_\pi$
from Definition \ref{def:4} and $\Upsilon_\pi$ from Definition
\ref{def:2} are close in variation distance. 

\begin{proposition}\label{PropThird}
  Let $\pi\in\mathcal P'_{\dickm \ell\cup\dickm r}$ and $\Xi_\pi$ and
  $\Upsilon_\pi$ be as in Definitions \ref{def:4} and \ref{def:2}.
  Then,
 \[ \sup_{\xi\in\mathcal P'_{\dickm \ell\cup\dickm r}} 
  \big|\mathbb P[\Xi_\pi = \xi] - \mathbb P[\Upsilon_\pi=\xi] \big| =
  \mathcal O\left( \frac{1}{(\log\alpha)^2}\right). \]
\end{proposition}

\begin{proof}
  We will only consider geometry (i). The proof for geometry (ii) is
  analogous. \\
  
  After realizing the splits in the beneficial background first
  according to the probabilities given in \eqref{Split} and
  \eqref{eq:Y2}, respectively, $\Xi_\pi$ and $\Upsilon_\pi$ are
  determined by the same equivalence relations \eqref{eq:equivGeoi}
  using the marks which hit the tree according to Definition
  \ref{def:4} and Table \ref{tab:marks}. Hence our proof consists of
  two steps. First, we show that the probabilities given in
  \eqref{Split} and \eqref{eq:Y2} differ only by
  $\mathcal{O}\left((\log \alpha)^{-2} \right)$. Second, we show that
  the error caused by generating the $SL$-, $LR$- and $SLR$-marks
  using \eqref{eq:Y3} instead of Definitions \ref{def:4} is
  $\mathcal{O}\left((\log \alpha)^{-2} \right)$.

  Both assertions rely on the same calculation. Assume a line in the
  Yule tree starts when the full Yule tree has $i_1$ lines for the
  last time and ends when the full Yule tree has $i_2>i_1$ lines for the
  last time.  Additionally, the line is hit by a Poisson process with
  rate $\rho = \gamma\frac{\alpha}{\log\alpha}$. The probability that
  the line is not hit by the Poisson process during the time the Yule
  process has $i$ lines, $i_1 < i\leq i_2$, is
  \[
  \frac{i \alpha}{i\alpha + \rho}
  \]
  because of competing exponential clocks. Analogously, the
  probability that the whole line is not hit, is, by a Taylor
  approximation,
  \begin{equation}\label{eq:rec5}\begin{aligned}
      \prod_{i=i_1+1}^{i_2} \frac{i\alpha}{i\alpha+\rho} &= \exp\left(
        \sum_{i=i_1+1}^{i_2} \log \left( 1 -
          \frac{\rho}{i\alpha+\rho}\right) \right) \\& = \exp\left(
        -\frac{\gamma}{\log\alpha} \sum_{i=i_1+1}^{i_2}
        \frac{1}{i+\rho/\alpha}\right) + \mathcal
      O\left(\frac{1}{(\log\alpha)^2}\right) \\ & = \exp\left(
        -\frac{\gamma}{\log\alpha} \sum_{i=i_1+1}^{i_2}
        \frac{1}{i}\right)+ \mathcal
      O\left(\frac{1}{(\log\alpha)^2}\right)= p_{i_1}^{i_2}(\gamma)+
      \mathcal O\left(\frac{1}{(\log\alpha)^2}\right),\end{aligned}
  \end{equation}
  since the neglected terms in the Taylor
  series are of order $\mathcal O \big( \rho^{2}/\alpha^2\big)=
  \mathcal O\big((\log \alpha)^{-2}\big)$ and higher.

  To prove that \eqref{Split} and \eqref{eq:Y2} coincide
  approximately, observe that 
 \[ \mathbb E\left[ \exp\left( - \rho\cdot \int_0^T X_s ds \right)\right] = 
  \mathbb E\left[ \exp\left( - \rho\cdot \int_0^T (1-X_s) ds \right)\right]
  \]
  by the time-reversibility of $\mathcal X$. Additionally, the right
  hand side gives the probability that a Poisson process with rate
  $\rho(1-X)$ does not hit a line by time $T$. By the random time
  change $d\tau = (1-X_t)dt$ this is approximately the same as the
  probability that a Poisson process with rate $\rho$ does not hit one
  line in a Yule tree until it has $\lfloor 2\alpha\rfloor$ lines and
  is hence given by $p_{0}^{\lfloor 2\alpha \rfloor}(\gamma)$.

  Next, we are considering the generation of the $SL$-, $LR$- and
  $SLR$-marks along the Yule tree. The probability that more than one
  event with rate $\rho_{SL}$ and $\rho_{LR}$ hits the Yule tree
  during the time it has $i$ lines is
  \[ \frac{\rho^2}{(i\alpha + \rho)^2} = \mathcal O\left(
    \frac{1}{(\log\alpha)^2}\right). \] Hence we can ignore this
  event. Together with the Markov property of the Poisson process we
  see that the marks on different lines in a sample tree may be
  generated independently once the topology and the total number of
  lines in the full Yule tree is known.

Consider a branch which starts when the full Yule tree has $i_1$ lines
and ends when it has $i_2$ lines. Using Definition \ref{def:4} this
line is hit by an $SL$-mark iff it is hit by the Poisson process at
rate $\rho_{SL}$ and an independent Poisson process with rate
$\rho_{LR}$ produces no mark between time $0$ and the time the Yule
tree has $i_2$ lines. Hence the probability for an $SL$-mark in
$\Xi_\pi$ is approximately given by
\begin{eqnarray}\begin{aligned} \nonumber
    &\left( 1 - \prod_{i=i_1+1}^{i_2}
      \frac{i\alpha}{i\alpha+\rho_{SL}}\right)\left( \prod_{i=1}^{i_2}
      \frac{i\alpha}{i\alpha+\rho_{LR}}\right) =
    \big(1-p_{i_{1}}^{i_{2}}(\gamma_{SL})\big)
    p_{0}^{i_{2}}(\gamma_{LR}) + \mathcal{O}\left(\frac{1}{(\log
        \alpha)^{2}}\right)
\end{aligned}\end{eqnarray}
If a branch is hit by the Poisson process with rate $\rho_{SL}$ but
did not obtain an $SL$-mark, it obtains an $SLR$-mark. Hence the
probability for such a mark is given by
\begin{eqnarray}\begin{aligned} \nonumber
    & \left( 1 - \prod_{i=i_1+1}^{i_2}
      \frac{i\alpha}{i\alpha+\rho_{SL}}\right)\left( 1
      -\prod_{i=1}^{i_2} \frac{i\alpha}{i\alpha+\rho_{LR}}\right) =
    \left(1-p_{i_{1}}^{i_{2}}(\gamma_{SL}\right)\left(1-
      p_{0}^{i_{2}}(\gamma_{LR})\right) +
    \mathcal{O}\left(\frac{1}{(\log \alpha)^{2}}\right)
\end{aligned}\end{eqnarray}
The branch is hit by an $LR$-mark if it is hit by the Poisson process
at rate $\rho_{LR}$ but not by the Poisson process with rate
$\rho_{SL}$. Hence the probability for an $LR$-mark is
\begin{eqnarray}\begin{aligned} \nonumber
    &\prod_{i=i_1+1}^{i_2}
    \frac{i\alpha}{i\alpha+\rho_{SL}}\left( 1 -\prod_{i=i_1+1}^{i_2}
    \frac{i\alpha}{i\alpha+\rho_{LR}}\right)
     = p_{i_{1}}^{i_{2}}(\gamma_{SL})\left(1- p_{i_{1}}^{i_{2}}(\gamma_{LR})\right) + \mathcal{O}\left(\frac{1}{(\log \alpha)^{2}}\right)
\end{aligned}\end{eqnarray}

As a consequence, the marks in $\mathbf Y_{|\pi'|}$ and
$\mathcal{Y}_{|\pi'|}$ coincide approximately (cf. Table
\ref{tab:marks}) and we are done.
 \end{proof}

\subsubsection*{Acknowledgement}
We thank Bernhard Haubold and Joachim Hermisson for comments on the
manuscript and Anton Wakolbinger and Franz Merkl for fruitful
discussion. We are grateful to Andy Lehnert, not only for help with
Figure \ref{sim}.

\newcommand{\etalchar}[1]{$^{#1}$}


\end{document}